\documentclass[aps,prd,preprintnumbers,showpacs,showkeys,nofootinbib,
superscriptaddress,fleqn,floatfix,tightenlines,10pt]{revtex4-1}
\usepackage{amsmath,amsfonts,amssymb,amscd,amsxtra,amsthm}
\usepackage{graphicx}  
\usepackage{epstopdf}
\usepackage{dcolumn}  
\usepackage{bm}          
\usepackage{slashed}
\usepackage{cancel}
\usepackage{float}
\usepackage{mathtools}
\usepackage{amsbsy} 
\usepackage{amstext}

\usepackage[utf8]{inputenc} 
\usepackage{booktabs}
\usepackage[normalem]{ulem} 
\usepackage[dvipsnames]{xcolor} 
\usepackage{tabularx}
\usepackage{enumitem}  
\usepackage{array} 
\usepackage{slashed}
\usepackage{tikz}
\usepackage{float}
\usepackage{multirow}
\renewcommand\sout{\bgroup \color{red} \ULdepth=-.5ex \ULset}

\makeatletter

\begin{document}  
\preprint{INHA-NTG-08/2019}
\title{Improved pion mean fields and masses of singly heavy baryons}  

\author{June-Young Kim}
\email[E-mail: ]{Jun-Young.Kim@ruhr-uni-bochum.de}
\affiliation{Ruhr-Universit\"at Bochum, Fakult\"at f\"ur Physik und
  Astronomie, Institut f\"ur Theoretische Physik II, D-44780 Bochum,
  Germany}
\affiliation{Department of Physics, Inha University, Incheon 22212,
Republic of Korea}

\author{Hyun-Chul Kim}
\email[E-mail: ]{hchkim@inha.ac.kr}
\affiliation{Department of Physics, Inha University, Incheon 22212,
Republic of Korea}
\affiliation{School of Physics, Korea Institute for Advanced Study 
  (KIAS), Seoul 02455, Republic of Korea}
\date{\today}
\begin{abstract}
A singly heavy baryon can be viewed as $N_c-1$ ($N_c$ as the number of
colors) light valence quarks bound by the pion mean fields that are
created by the presence of the $N_c-1$ valence quarks
self-consistently, while the heavy quark inside a singly heavy baryon
is regarded as a static color source. We investigate how the pion mean
fields are created by the presence of $N_c$, $N_c-1$, and $N_c-2$
light valence quarks, which correspond to the systems of light
baryons, singly heavy baryons, and doubly heavy baryons. As the number
of color decreases from $N_c$ to $N_c-1$, the pion mean fields undergo
changes. As a result, the valence-quark contributions to the moments
of inertia of the soliton become larger than the case of the $N_c$
valence quarks,  whereas the sea-quark contributions decrease
systematically. On the other hand, the presence of the $N_c-2$ valence
quarks is not enough to produce the strong pion mean fields, which
leads to the fact that the classical soliton can not be formed. It
indicates that the pion mean-field approach is not suitable to
describe doubly heavy baryons. We show that the mass spectra of the
singly heavy baryons are better described by the improved pion mean
fields, compared with the previous work in which the pion mean fields
are assumed to be intact with $N_c$ varied. 
\end{abstract}
\maketitle
\section{Introduction}
A light baryon can be regarded as a state of $N_c$ (the number
of colors) valence quarks bound by meson mean fields in the large
$N_c$ quantum chromodynamics (QCD)~\cite{Witten:1979kh, Witten:1983, 
  Witten:1983tx}. Since the nucleon mass is proportional to $N_c$
whereas meson-loop fluctuations are suppressed by $1/N_c$,  the
mean-field approach is justified in the large $N_c$ limit.
The chiral quark-soliton model ($\chi$QSM) realizes effectively 
this idea of the pion mean-field approach~\cite{Diakonov:1987ty,
  Christov:1995vm, Diakonov:1997sj,Blotz:1992pw}.   
The model has been very successful in describing the lowest-lying
SU(3) light baryons. The same idea can be applied to a singly heavy
baryon, which consists of two light valence quarks and one heavy
quark. We can consider the heavy baryon as the $N_c-1$ light valence
quarks bound by the pion mean fields~\cite{Diakonov:2010tf}. In the
limit of infinitely heavy quark mass ($m_Q\to \infty $), the heavy
quark can be regarded as a static color source. 
This pion mean-field approach has a great virtue because it describes
both the light and singly heavy baryons on an equal footing. Recently, it
was shown that the pion mean-field approach 
indeed describes very well the masses of the singly heavy
baryons~\cite{Yang:2016qdz, Kim:2018xlc}. The magnetic moments and 
electromagnetic form factors of the singly heavy baryons were also
studied within this approach with all the parameters fixed in the light baryon
sector~\cite{Yang:2018uoj, Kim:2018nqf}.
Very recently, the LHCb Collaboration announced the five or six
excited $\Omega_c$'s, among which two of them have unusually small 
widths~\cite{Aaij:2017nav}. The Belle Collaboration confirmed the
existence of the four excited $\Omega_c$'s~\cite{Yelton:2018mag}. 
In the $\chi$QSM, the two of the newly found excited $\Omega_c$'s were
classified as the members of the baryon antidecapentaplet where
remaining $\Omega_c$'s belong to the excited sextet
representations~\cite{Kim:2017jpx}.  The small widths of those two 
$\Omega_c$'s were well explained in the $\chi$QSM~\cite{Kim:2017khv}.

One can ask a critical question about the large $N_c$ limit, since
$N_c$ and $N_c-1$ are parametrically not different when the limit
of $N_c\to \infty$ is taken. In fact, the large $N_c$ limit is
introduced to justify the mean-field approach in which the
$1/N_c$-order meson fluctuations or in a more traditional language,
particle-hole excitations, can be neglected. When it comes to
the real world, i.e., when one sets $N_c=3$, certain important physics
of baryons attributed to the large $N_c$ limit is still inherited. For
example, the pion mean-field approach at $N_c=3$ describes very well
various properties and observables of the lowest-lying SU(3)
baryons~(see for example a review~\cite{Christov:1995vm}). The same is
true if one takes $N_c-1=2$ as mentioned above. While $N_c$ and
$N_c-1$ are parametrically the same in the large $N_c$ limit, the real
world at $N_c=3$ exhibits certain difference between the light and
singly heavy baryons. For example, the $N_c=3$ chiral soliton consists
of three quarks such that the soliton is fermionic, whereas the
$N_c-1=2$ soliton emerges as a colored bosonic soliton in the
antisymmetric color state. It yields a singly-heavy baryon in the
color singlet state when it is coupled to a heavy quark. Moreover, the
hypercharge $Y'=(N_c-1)/3=2/3$ describes very well the SU(3)
representations of the singly heavy baryons including excited ones
(see for example, a recent review~\cite{Kim:2018cxv}). Thus, we will
consider the large $N_c$ limit in this work to justify the existence
of the pion mean-field solution rather than a mathematically rigorous
limit. 

Previous works~\cite{Yang:2016qdz, Kim:2017jpx, Kim:2017khv}
employed an ``model-independent approach'', which means that all the
dynamical parameters were fixed by using the experimental data. While
this approach has a merit to predict the experimental data without any
model calculations, one can not decompose the valence- and sea-quark
contributions, so that the overall replacement of the $N_c$ factor by
$N_c-1$ underestimates the sea-quark contributions. Thus, it is
inevitable to introduce an additional parameter to compensate it.
On the other hand, the self-consistent $\chi$QSM, where the pion mean
fields are created explicitly by solving the classical equation of
motion, assumed that the pion mean fields are not modified by changing
the number of the valence quarks~\cite{Kim:2018nqf}. In
Ref.~\cite{Kim:2018nqf}, the number of the valence quarks $N_c$ are
merely replaced by $N_c-1$ to describe the singly heavy baryons with
the same pion mean-field solutions used. However, we find that the
number of the valence quarks indeed alter the pion mean fields, which
will be shown in the present work. We expect that the reduction of the
number of the valence quarks from $N_c$ to $N_c-1$ will create weaker
vacuum polarizations and as a result will lead to the weaker pion mean
fields. In this work, we will explicitly compute the classical
equations of motion to derive the pion mean-field solution, changing
the number of the valence quarks. Interestingly, we find that the pion
mean-field solutions do not exist when the number of the valence
quarks is $N_c-2$. This is understandable, since $N_c-2$ means
practically a single light valence quark. The presence of a single
valence quark is not enough to create a strong pion mean fields to
bind a doubly heavy baryon. Thus, in any pion mean-field approaches,
we are not able to describe a system of doubly heavy baryons. In the
present work, we will revisit the mass splittings of the singly heavy
baryons including the baryon antitriplet, sextet, and
antidecapentaplet. We also want to mention that the modification of the
pion mean fields in the presence of the $N_c-1$ valence quarks has
another important physical implications. A recent work on the
gravitational form factors of the singly heavy baryons indicate that
the stability condition or the von Laue
condition~\cite{Polyakov:2018zvc} for the singly heavy baryons can
only be satisfied by using the present modified pion mean
fields~\cite{Kim}.  

The present paper is organized as follows: In Section~\ref{sec:2}, we
briefly explain the $\chi$QSM. Starting from the baryon correlation
function, we show how the pion mean-field solution can be obtained.
Then we introduce the collective zero-mode quantization of the
baryon and derive the collective Hamiltonian. The collective wave
functions for the singly heavy baryons are obtained by diagonalizing
the Hamiltonian and coupling the SU(3) wave functions to the heavy
quark. In Section~\ref{sec:3}, we first discuss the results of the
pion mean-field solutions or the soliton profile functions. We then
present the results of the classical masses as functions of the
dynamical quark mass and discuss how the pion mean fields influence
them. We also show explicitly that the pion mean-field solution does
not exist when the number of the valence quarks $N_c-2$. Finally, we
present the results of the masses of the singly heavy baryons
including the baryon antitriplet, sextet, and antidecapentaplet. 
In the final Section, we summarize the present work and draw
conclusions. 

\section{Heavy baryons in the chiral quark-soliton
  model \label{sec:2}} 
\subsection{Nucleon correlation function}
In the limit of the infinite heavy-quark mass ($m_Q\to \infty$), 
heavy quarks inside a singly or doubly heavy baryon can be viewed as a
static color source. It means that the heavy quarks play a mere role
to make the heavy baryon a color singlet. In order to describe the
heavy baryons within the $\chi$QSM, we consider the baryon correlation
functions consisting of $N_c-N_Q$ light valence quarks in Euclidean
space, where $N_Q$ ($N_Q\le 2$) denotes the number of the heavy quarks
involved. This is plausible, since the heavy-quark 
propagators in the limit of $m_Q\to \infty$ contribute to the
correlation function of the singly or doubly heavy baryon only in a
trivial way. Thus, the correlation function of the singly or doubly
heavy baryon can be expressed as 
\begin{align}
\Pi_{B}(\bm{x}-\bm{y},T)  &= \langle \mathcal{J}_B (\bm{x}, T/2)
    \mathcal{J}_B^\dagger  (\bm{y},-T/2) \rangle_0 \cr
    &= \frac{1}{\mathcal{Z}}\int  \mathcal{D} U
  \mathcal{D}\psi^\dagger 
  \mathcal{D}\psi \mathcal{J}_B(0,T/2) \mathcal{J}_B^\dagger
  (0,-T/2) 
  e^{\int d^4 x\,\psi^\dagger (i\rlap{/}{\partial} + i
  MU^{\gamma_5}+ i \hat{m})\psi},
\label{eq:corr1}
\end{align} 
where $\mathcal{Z}$ is the low-energy effective chiral partition function. 
$\mathcal{J}_B$ represents the Ioffe-type baryonic current that
consists of $N_c-N_{Q}$ light valence quarks for a singly or doubly heavy
baryon $B$  
\begin{align}
\mathcal{J}_B(x) &= \frac1{(N_c-N_{Q})!}
  \varepsilon^{\alpha_1\cdots\alpha_{N_c-N_{Q}}}
                   \Gamma_{J'J'_3,TT_3}^{\{f_1\cdots f_{N_c-N_Q}\}}
  \psi_{\alpha_1 f_1}(x) \cdots \psi_{\alpha_{N_c-N_{Q}}f_{N_c-N_{Q}}}
  (x),  
\end{align}
$\alpha_i$ denote color indices. $\Gamma_{J'J'_3,TT_3}^{f}$ is a
symmetric matrix with flavor and spin indices $f$. $J'$ and $T$
represent the spin and isospin of the heavy baryon, respectively and
$J'_3$ and $T_3$ are the corresponding third components of them,
respectively. The notation $\langle \cdots \rangle_0$ in
\eqref{eq:corr1} stands for the vacuum expectation value. $M$
designates the dynamical quark mass and the chiral 
field $U^{\gamma_5}$ is defined as
\begin{align}
U^{\gamma_5} = U\frac{1+\gamma_5}{2} + U^\dagger \frac{1-\gamma_5}{2},
\end{align}
with 
\begin{align}
U = \exp\left[i\frac{\pi^a \lambda^a}{f_\pi}\right].  
\end{align}
$\pi^a$ represents the pseudo-Nambu-Goldstone(NG) field.
$\hat{m}$ is the mass matrix of the current quarks, which is written
as $\hat{m} = \mathrm{diag}(m_{\mathrm{u}}, \, m_{\mathrm{d}}, \,
m_{\mathrm{s}})$. We assume in the present work isospin symmetry,
i.e. $m_{\mathrm{u}}=m_{\mathrm{d}}$. Thus, we introduce the average 
mass of the up and down quarks $m_0 = (m_{\mathrm{u}} +
m_{\mathrm{d}})/2$. The strange current quark mass $m_{\mathrm{s}}$
will be treated perturbatively to linear order.   
Since we introduce hedgehog ansatz or hedgehog symmetry, we consider 
the trivial embedding~\cite{Witten:1983}
\begin{align}
U(\bm{r}) = \begin{pmatrix}
U_{\mathrm{SU(2)}}(\bm{r})& 0 \\
0 & 1
\end{pmatrix},
\label{eq:embed}
\end{align}
where $\bm{n}$ is defined as the normalized radial vector $\bm{n}=
\bm{r}/r$, and $U_{SU(2)}=\exp[i \bm{n}\cdot \bm{\tau}
\Theta(r)]$. $\bm{\tau}$ are the Pauli matrices in isospin
space. $\Theta(r)$ denotes the profile function of the soliton, which
will be obtained in a self-consistent way by the minimizing procedure.

Integrating over the quark fields, we obtain the following expression
of the baryonic correlation function as 
\begin{align}
\Pi_{B}(\bm{x}-\bm{y}, T) =
  \frac{1}{Z}\Gamma_{J'J'_3,TT_3}^{\{f\}} \Gamma_{J'J'_3,TT_3}^{\{g\}*} \int
  \mathcal{D} U \prod_{i=1}^{N_c-N_{Q}}  \left\langle
  \bm{x},T/2,\alpha_i\left|   \frac1{D(U)} \right|\bm{y},-T/2,\beta_i
  \right\rangle   e^{-S_{\mathrm{eff}}(U)}, 
\label{eq:corr2}
\end{align}  
where the one-body Dirac operator $D(U)$ is defined by 
\begin{align}
D(U) = i\gamma_4 \partial_4 + i\gamma_k \partial_k + i MU^{\gamma_5} +
  i \hat{m}.
\end{align}
$S_{\mathrm{eff}}$ represents the effective chiral action written
as 
\begin{align}
S_{\mathrm{eff}} = -N_c \mathrm{Tr}\log D(U).  
\label{eq:effecXac}
\end{align}
Here, $\mathrm{Tr}$ denotes the functional trace over space-time and
all internal spaces. 
Taking the Euclidean time to be infinity ($T\to \infty$), we can pick  
up lowest-lying baryon states from the correlation 
function~\cite{Diakonov:1987ty, Christov:1995vm} as 
\begin{align}
 \Pi_B(\bm{x}-\bm{y}, T) \sim \exp[- (N_c-N_{Q}) E_{\mathrm{val}} +
  E_{\mathrm{sea}} T],   
\end{align}
where $E_{\mathrm{val}}$ and $E_{\mathrm{sea}}$ the valence and
sea quark energies. However, the mean fields $U$ being involved in the
calculation should be determined.  Note that the profile function $
\Theta(r)$ satisfies the boundary conditions at two end points,
i.e. $\Theta(0) =\pi$ and $\Theta(\infty) = 0$. The SU(2)
single-quark Hamiltonian $h_{\mathrm{SU(2)}}(U)$ is 
defined as  
\begin{align}
h_{\mathrm{SU(2)}}(U) = i \gamma_{4}\gamma_{i} \partial_{i} -
  \gamma_{4} M U_{\mathrm{SU(2)}}^{\gamma^{5}} - \gamma_{4} m_0 . 
\end{align}
Then, the one-body Dirac equation is written as 
\begin{align}
h_{\mathrm{SU(2)}}(U) \Phi_{n}(\bm{r}) = E_{n} \Phi_{n}(\bm{r}),
\label{eq:one_enregy}
\end{align}
where $E_n$ denote the eigen-energies of the one-body Hamiltonian
$h_{\mathrm{SU(2)}}(U)$. 

\subsection{Classical equation of motion and self-consistent solution} 
The classical equation of motion can be derived by minimizing the
energy of the classical soliton 
\begin{align}
\left.\frac{\delta}{\delta \Theta(\bm{r})}[ (N_c-N_{Q})
  E_{\mathrm{val}} + 
  E_{\mathrm{sea}}]\right|_{\Theta_c} = 0, 
\label{eq:saddle}
\end{align}
where $\Theta_c$ is the soliton profile function at the stationary
point or the pion mean-field solution. Solving Eq.~\eqref{eq:saddle},
we find the equation of motion  
\begin{align}
\sin \Theta(\bm{r}) S(\bm{r}) - \cos \Theta(\bm{r}) P(\bm{r}) = 0,   
\label{eq:eq_motion}
\end{align}
where $S(\bm{r})$ and $P(\bm{r})$ are defined as 
\begin{align}
S(\bm{r})  &=  M \left[  N_{c} \sum_{n} R^{\Lambda}_{2}(E_{n})
             \overline{\Phi}_{n}(\bm{r}){\Phi}_{n}(\bm{r})
             +{(N_{c}-N_{Q})}\theta(E_{\mathrm{val}})
             \overline{\Phi}_{\mathrm{val}}(\bm{r}){\Phi}_{\mathrm{val}}
             (\bm{r})\right],
             \cr 
P(\bm{r}) &=  M \left[  N_{c} \sum_{n} R^{\Lambda}_{2}(E_{n})
            \overline{\Phi}_{n}(\bm{r}) i
            \gamma_{5}(\tau^{i}{n}^{i}){\Phi}_{n}(\bm{r})
            +{(N_{c}-N_{Q})}\theta(E_{\mathrm{val}})
            \overline{\Phi}_{\mathrm{val}}(\bm{r})i
            \gamma_{5}(\tau^{i}{n}^{i}){\Phi}_{\mathrm{val}}(\bm{r})\right]. 
\label{eq:densities}
\end{align}
Here,  $\Phi_{\mathrm{val}}(\bm{r})
=\langle \bm{r}|\mathrm{val}\rangle$ and $ \Phi_{n}(\bm{r}) =\langle
\bm{r} |n\rangle$ denote the single-particle wave functions of the  
valence and sea quarks with the corresponding eigen-energies
$E_{\mathrm{val}}$ and $E_n$ of the one-body Hamiltonian
$h_{\mathrm{SU}(2)}(U_c)$, respectively. The regularization function
$R^{\Lambda}_{2}(E_{n})$ is defined as 
\begin{align} 
R^{\Lambda}_{2}(E_{n}) = \frac{1}{4\sqrt{\pi}} \int \phi(u) \frac{d
  u}{u^{1/2}} E_{n} e^{-u E_{n}^{2}}, 
\end{align}
where $\phi(u)=c\theta(u-\Lambda_{1}^{-2}) +
(1-c)\theta(u-\Lambda_{2}^{-2})$~\cite{Blotz:1992pw}. The values of
the parameters are taken from Ref.~\cite{Kim:2018xlc}. 
Thus, the soliton mass is finally derived as 
\begin{align}
M_{\mathrm{sol}} = (N_c-N_{Q}) \theta(E_{\mathrm{val}})
  E_{\mathrm{val}}(U_c) + E_{\mathrm{sea}}(U_c).   
\label{eq:solnc}
\end{align}
When $N_Q=0$, the result of $M_{\mathrm{sol}}$ is the same as in the
original $\chi$QSM for the light baryons. However, when $N_Q=1$, we
obtain the modified pion mean-field solutions for the singly heavy 
baryons. Note that in Ref.~\cite{Kim:2018xlc} the pion mean fields are
assumed to be not changed. In the present work, we will show
explicitly that it was not a correct assumption. As for the case of
$N_Q=2$, we find that there are no pion mean-field solutions. It
implies that the present scheme of the mean-field approach does not
apply to the description of doubly heavy baryons. When the dynamical
quark mass is almost two times larger than its usual value ($M\simeq
400$ MeV), we can find the solution of Eq.~\eqref{eq:eq_motion}.   
When $N_Q\neq 0$, the classical mass is changed to be 
\begin{align}
M_{\mathrm{cl}} = M_{\mathrm{sol}} + N_{Q} m_Q,  
\label{eq:classical_mass}
\end{align}
where $m_Q$ is the \textit{effective} heavy quark mass that contains
also the binding energy of the heavy quark. Thus, it is different from
that discussed in QCD and will be absorbed in the center mass of each
representation, which will be discussed later. Note that when the
level of the valence quarks crosses the line where the valence
energy becomes negative, the soliton mass is given solely by the
sea-quark energy. 

\subsection{Zero-mode collective quantization}
Having carried out the zero-mode quantization~\cite{Christov:1995vm},
we arrive at the collective Hamiltonian for singly heavy 
baryons 
\begin{align}
H =& H_{\mathrm{sym}} + H^{(1)}_{\mathrm{sb}},
\label{eq:Hamiltonian}
\end{align}
where $H_{\mathrm{sym}}$ represents the flavor SU(3) symmetric part
\begin{align} 
H_{\mathrm{sym}}=M_{\mathrm{cl}}+\frac{1}{2I_{1}}\sum_{i=1}^{3}
\hat{J}^{2}_{i} +\frac{1}{2I_{2}}\sum_{a=4}^{7}\hat{J}^{2}_{a}.  
\label{eq:sym}
\end{align} 
Here, $I_{1}$ and $I_{2}$ denote the moments of inertia of the
soliton. The explicit expressions for $I_{1,\,2}$ are given in
Ref.~\cite{Blotz:1992pw, Christov:1995vm}. 
Note that the second and third terms in Eq.~\eqref{eq:sym} arise from
the rotation of the chiral soliton, which is of order $1/N_c$
($I_{1,2}\sim N_c$).  
The operators
$\hat{J}_{i}$ and $\hat{J}_{a}$ represent the spin generators in
SU(3). In the $(p,\,q)$ representation of the SU(3) group, we find the
eigenvalue of the SU(3) quadratic Casimir operator $\sum_{i=1}^8 
J_i^2$ as    
\begin{align}
C_2(p,\,q) = \frac13 \left[p^2 +q^2 + pq + 3(p+q)\right].   
\end{align}
Thus, the eigenvalues of $H_{\mathrm{sym}}$ are obtained as  
\begin{align} 
E_{\mathrm{sym}}(p,q) = M_{\mathrm{cl}}+ \frac{1}{2I_{1}} J(J+1) 
+\frac{1}{2I_{2}}\left[C_2(p,\,q) - J(J+1)\right] 
-\frac{3}{8I_{2}} Y'^2.
\label{eq:RotEn}
\end{align} 
The right hypercharge $Y'$ is constrained to be $(N_c-N_Q)/3$,
which is imposed by the $N_c-N_Q$ valence quarks inside a singly heavy
baryon. Thus, $Y'$ counts effectively the number of the valence quarks
involved. Note that in the Skyrme model the right hypercharge is
constrained by the Wess-Zumino term. When $Y'=1$ with $N_Q=0$,
i.e. when the light baryons are concerned, it provides the selection
rule. That is, only the baryon representations that contain $Y=1$ are
allowed such as the baryon octet ($\bm{8}$), decuplet ($\bm{10}$),
antidecuplet ($\overline{\bm{10}}$), eikosiheptaplet ($\bm{27}$),
etc., all of which contain the baryons with $Y=1$. When it comes to
the singly heavy baryons, the right hypercharge becomes
$Y'=2/3$. Thus, allowed representations are the baryon antitriplet
($\overline{\bm{3}}$), sextet ($\bm{6}$) with $J=1/2$ and $J=3/2$,
antidecapentaplet ($\overline{\bm{15}}$) with $J=1/2$ and $J=3/2$,
etc., which include the singly heavy baryons with $Y=2/3$.

The collective wavefunctions of the baryons are derived as 
\begin{align} 
\psi_B^{({\mathcal{R}})}(J'J'_3,J;A)=
\sqrt{\mathrm{dim}(p,\,q)} (-1)^{-\frac{ \overline{Y}  }{2}+J_3}
  D^{(\mathcal{R})\ast}_{(Y,T,T_3)(\overline{Y}  ,J,-J_3)}(A),  
\label{eq:waveftn}
\end{align} 
where 
\begin{align}
\mathrm{dim}(p,\,q) = (p+1)(q+1)\left(1+\frac{p+q}{2}\right).  
\end{align}
$J$ stands for the soliton spin, and ${J_{3}}$ represents
its third component, respectively. 

\subsection{Collective Hamiltonian for flavor SU(3) symmetry breaking} 
The symmetry-breaking part of the collective
Hamiltonian is given as~\cite{Blotz:1992pw, Christov:1995vm}
\begin{align} 
H^{(1)}_{\mathrm{sb}} 
=&\frac{\Sigma_{\pi N}}{m_{0}}\frac{m_{\mathrm{s}}}{3}
+\alpha D^{(8)}_{88}+ \beta \hat{Y}
+ \frac{\gamma}{\sqrt{3}}\sum_{i=1}^{3}D^{(8)}_{8i}
\hat{J}_{i},
\label{eq:sb}
\end{align}
where
\begin{align} 
\alpha=\left (-\frac{\Sigma_{\pi N}}{3m_0}+\frac{
  K_{2}}{I_{2}} Y  
\right )m_{\mathrm{s}},
 \;\;\;  \beta=-\frac{ K_{2}}{I_{2}}m_{\mathrm{s}}, 
\;\;\;  \gamma=2\left ( \frac{K_{1}}{I_{1}}-\frac{K_{2}}{I_{2}} 
 \right ) m_{\mathrm{s}}.
\label{eq:alphaetc}
\end{align}
The first term in Eq.~\eqref{eq:sb} can be absorbed into the symmetric
part of the Hamiltonian, since it does not contribute to the mass
 splittings of the baryons in a given representation. 
The three parameters $\alpha$, $\beta$, and $\gamma$ are
expressed in terms of the moments of inertia $I_{1,\,2}$ and
$K_{1,\,2}$. Since we have $N_c-N_Q$ light quarks in the case of the 
heavy baryons, we need to modify the valence contributions to the
moments of inertia and the sigma $\pi N$ term. The modification can be
done easily by replacing the prefactor $N_c$, which counts the number
of the valence quarks, by $N_c-N_Q$. The explicit expressions for the
moments of inertia and the $\pi N$ sigma term can be found in
Ref.~\cite{Kim:2018xlc}. 

The effects of flavor SU(3) symmetry breaking being introduced, the 
collective wavefunctions are no longer expressed by a pure
representation but are mixed with other representations. Dealing with
the collective Hamiltonian~\eqref{eq:sb} as a small perturbation and
using the second-order perturbation theory, we obtain the
wavefunctions for the baryon anti-triplet ($J=0$) and the sextet
($J=1$) respectively~\cite{Kim:2018xlc} as 
\begin{align}
 & |B_{\overline{\bm{3}}_{0}}\rangle
=|\overline{\bm{3}}_{0},B\rangle+p_{\overline{15}}^{B}|
   \overline{\bm{15}}_{0},B\rangle,  
\cr
 & |B_{\bm{6}_{1}}\rangle
=|{\bm{6}}_{1},B\rangle+q_{\overline{15}}^{B}
|{\overline{\bm{15}}}_{1},B\rangle+q_{\overline{24}}^{B}|
{{\overline{\bm{24}}}_{1}},B\rangle,
\cr
\label{eq:mixedWF1}
\end{align}
with the mixing coefficients 
\begin{align}
p_{\overline{15}}^{B}\;\;=\;\;p_{\overline{15}}
\left[
\begin{array}{c}
-\sqrt{15}/10\\
-3\sqrt{5}/20
\end{array}
\right], 
\hspace{3em}
q_{\overline{15}}^{B}\;\;=\;\;q_{\overline{15}}
\left[
\begin{array}{c}
\sqrt{5}/5\\
\sqrt{30}/20\\
0
\end{array}
\right], 
\hspace{3em}
q_{\overline{24}}^{B}\;\;=\;\;q_{\overline{24}}
\left[
\begin{array}{c}
-\sqrt{10}/10\\
-\sqrt{15}/10\\
-\sqrt{15}/10
\end{array}
\right],
\label{eq:pqmix-1}
\end{align}
respectively, in the basis $\left[\Lambda_{Q},\;\Xi_{Q}\right]$ for
the anti-triplet and
$\left[\Sigma_{Q}\left(\Sigma_{Q}^{\ast}\right),
\;\Xi_{Q}^{\prime}\left(\Xi_{Q}^{\ast}\right),
\;\Omega_{Q}\left(\Omega_{Q}^{\ast}\right)\right]$
for the sextets. The expressions for the parameters
$p_{\overline{15}}$, $q_{\overline{15}}$, and $q_{\overline{24}}$ are
also found in Refs.~\cite{Yang:2018uoj, Kim:2018xlc}.
Note that the mixing coefficients are proportional to $m_{\mathrm{s}}$
linearly.   

The complete wavefunction for a heavy baryon can be constructed by
coupling the soliton wavefunction to the heavy quark such that the
heavy baryon becomes a color singlet, which is expressed as 
\begin{align}
|B_\mu; (J',J_3')(T,T_3) \rangle 
=\sum_{J_{3},\,J_{Q3}}C_{\,J,J_{3}\,J_{Q}\,J_{Q3}}^{J'\,J_{3}'}
  \;\mathbf{\chi}_{J_{Q3}}\; 
|B_{\mu}; (J,J_3)(T,T_3)\rangle.
\label{eq:HeavyWF-1}
\end{align}
Here, $\mathbf{\chi}_{J_{Q3}}$ denotes the heavy-quark spinor and
$C_{\,J,J_{3}\,J_{Q}\,J_{Q3}}^{J'\,J_{3}'}$ represent the
corresponding Clebsch-Gordan coefficients. $|B_{\mu};
(J,J_3)(T,T_3)\rangle$ means the collective wavefunctions of the
quantized light soliton, given already in Eq.~\eqref{eq:waveftn}.

\subsection{Baryon antitriplet and sextet}
Taking into account the $m_{\mathrm{s}}$ corrections to the first
order, we can write the masses of the singly heavy baryons in
representation $\mathcal{R}$ as  
\begin{align}
M_{B,\mathcal{R}}^Q = M_{\mathcal{R}}^Q + M_{B,\mathcal{R}}^{(1)},  
\label{eq:FirstOrderMass}
\end{align}
where 
\begin{align}
M_{\mathcal{R}}^Q = M_{\mathrm{cl}} + E_{\mathrm{sym}}(p,q).  
\label{eq:center_mass}
\end{align}
$M_{\mathcal{R}}^Q$ is called the center mass of a heavy baryon in
representation $\mathcal{R}$. $E_{\mathrm{sym}}(p,q)$ is defined in
Eq.~\eqref{eq:RotEn}. The lower index $B$ denotes a certain
baryon belonging to a specific representation $\mathcal{R}$. The upper
index $Q$ stands for either the charm sector ($Q=c$) or the bottom
sector ($Q=b$). The center masses for the anti-triplet and sextet
representations can be explicitly written as
\begin{align}
\label{eq:center_mass1}
M_{\overline{\bm{3}}}^Q = M_{\mathrm{cl}} + \frac1{2I_2}, \;\;\; 
M_{\bm{6}}^Q = M_{\overline{\bm{3}}}^Q +  \frac1{I_1},
\end{align}
where $M_{\mathrm{cl}}$ was defined in Eq.~\eqref{eq:classical_mass}.
Note that the mass splitting between the baryon anti-triplet and sextet
are determined by $1/I_{1}=178$~MeV in the model, while the experimental 
data~\cite{Yang:2016qdz,Tanabashi:2018oca} suggest $1/I_1 \approx
172$MeV. The second term in Eq.~(\ref{eq:FirstOrderMass}) denotes the 
linear-order $m_{\mathrm{s}}$ corrections to the heavy baryon mass 
\begin{align}
M^{(1)}_{B,{\cal{R}}} = \langle B, {\cal{R}} | H_{\mathrm{sb}}^{(1)} 
| B, {\cal{R}} \rangle  = Y\delta_{{\cal{R}}},
\end{align}
where
 \begin{align} 
&\delta_{\overline{\bm{3}}}=\frac{3}{8}\alpha+\beta, \;\;\;\;
\delta_{\bm{6}}=\frac{3}{20}\alpha+\beta-\frac{3}{10}\gamma. 
\label{eq:delta36}
\end{align}
The values of the matrix elements for the relevant SU(3) Wigner $D$
functions are tabulated in Ref.~\cite{Kim:2018xlc}.
Thus, we obtain the masses of the lowest-lying singly heavy baryons as 
\begin{align}
M_{B,\overline{\bm{3}}}^Q = M_{\overline{\bm{3}}}^Q  +
                     Y \delta_{\overline{\bm{3}}} ,\;\;\;
M_{B,\bm{6}}^Q =M_{\bm{6}}^Q  + Y  \delta_{\bm{6}}, 
  \label{eq:firstms}
\end{align}
with the linear-order $m_{\mathrm{s}}$ corrections taken into account. 

While the baryon sextet with $J'=1/2$ and $J'=3/2$ are degenerate in
the limit of the infinitely heavy-quark mass ($m_Q\to \infty$), the
degeneracy is removed in reality. Thus, we need to introduce  
the hyperfine interaction that will lift the degeneracy 
of different spin states in the sextet representation. This
interaction is introduced phenomenologically. So, we will fix
the hyperfine interactions by using the experimental data, as was
proposed by Ref.~\cite{Yang:2016qdz}. The spin-spin interaction
Hamiltonian is written as  
\begin{align}
H_{solQ} = \frac23\frac{\kappa}{m_Q M_{\mathrm{sol}}} \bm{J}\cdot
   \bm{J}_Q = \frac23 \frac{\varkappa}{m_Q} \bm{J}\cdot
   \bm{J}_Q,
 \label{eq:hyperf_H}
\end{align}
where $\kappa$ represents the flavor-independent hyperfine coupling
constant. Note that the baryon anti-triplet does not acquire any
contribution from the hyperfine interaction, since the corresponding
soliton has spin $J=0$. On the other hand, the baryon sextet has
$J=1$. Being coupled to the heavy quark spin, it produces two
different multiplets, i.e., $J'=1/2$ and $J'=3/2$, of which the masses
are expressed respectively as  
\begin{align}
  \label{eq:hyperf_Masses}
 M_{B,\bm{6}_{1/2}}^Q = M_{B,\bm{6}}^Q -\frac23
  \frac{\varkappa}{m_Q},\;\;\;  
M_{B,\bm{6}_{3/2}}^Q =   M_{B,\bm{6}}^Q +\frac13
  \frac{\varkappa}{m_Q}.
\end{align}
Thus, we find the hyperfine mass splitting as  
\begin{align} 
M_{B,\bm{6}_{3/2}}^Q - M_{B,\bm{6}_{1/2}}^Q  =
  \frac{\varkappa}{m_Q}, 
\end{align}
where the corresponding numerical value can be determined by using the
center value of the sextet masses. In the charmed and bottom baryon 
sectors, we obtain the corresponding numerical values respectively  
\begin{align}
\frac{\varkappa}{m_c} =  68.1\,  \mathrm{MeV}, \;\;\;
\frac{\varkappa}{m_b} = 20.3\,  \mathrm{MeV},  
\label{eq:hf_result}
\end{align}
which were already shown in Ref.~~\cite{Yang:2016qdz}.

\subsection{Heavy pentaquarks}
The five resonances of the $\Omega_{c}$ were reported by the
LHCb~\cite{Aaij:2017nav} Collaboration, of which the four $\Omega_c$'s
were confirmed by the Belle~\cite{Yelton:2018mag}
Collaboration. In a recent work~\cite{Kim:2017jpx}, 
$\Omega_{c}$~(3050) and $\Omega_{c}$~(3119) with very small decay
width among the five $\Omega_c$'s were identified as heavy pentaquarks
belonging to the baryon antidecapentaplet. The
baryon $\overline{\bm{15}}$ was originally proposed by
Diakonov~\cite{Diakonov:2010tf}, since it appears naturally from the
$\chi$QSM with $Y=2/3$. In the present work, we compute the mass
splittings of the baryon antidecapentaplet self-consistently. 
Being similar to the baryon sextet, the antidecapentaplet consists of
the two representations with spin $J'=1/2$ and $J'=3/2$.

The expression of the mass formulae of the heavy pentaquarks comes from
Eq.~\eqref{eq:FirstOrderMass}. The center-mass formulae of the
antidecapentaplet are expressed as 
\begin{align}
\label{eq:center_mass2}
&M_{\overline{\bm{15}},J=0}^Q = M_{\mathrm{cl}} +
  \frac{5}{2}\frac{1}{I_2} = M_{\overline{\bm{3}}}^Q+\frac{2}{I_{2}} ,
  \cr 
&M_{\overline{\bm{15}},J=1}^Q = M_{\mathrm{cl}} +
  \frac{3}{2}\frac{1}{I_2}+ \frac{1}{I_1} =
  M_{{\bm{6}}}^Q+\frac{1}{I_{2}}, 
\end{align}
The value of the moment of inertia $I_{2}$ will be explicitly
given in Table~\ref{tab:1} later, i.e. $1/I_{2}=379$ MeV. Note that
its value is smaller than with that of $1/I_{2}=400-450$ MeV, which
was obtained in Ref.~\cite{Kim:2017jpx}. 

As mentioned previously, there are two baryon $\overline{\bm{15}}$
representations with spin $J'=1/2$ and $J'=3/2$, which are degenerate
in the limit of $m_Q\to \infty$. Thus, we need to introduce the same
hyperfine interaction given in Eq.~\eqref{eq:hyperf_H} to remove the
degeneracy. The explicit expressions of the center masses are then
given as  
\begin{align}
&M_{B,\overline{\bm{15}}_{1/2},J=0}^Q = M_{\overline{\bm{3}}}^Q +
  \frac{2}{I_{2}}, \cr 
&M_{B,\overline{\bm{15}}_{1/2},J=1}^Q = M_{{\bm{6}}}^Q +
  \frac{1}{I_{2}} -\frac23 \frac{\varkappa}{m_Q},  \cr 
&M_{B,\overline{\bm{15}}_{3/2},J=1}^Q = M_{{\bm{6}}}^Q +
  \frac{1}{I_{2}} +\frac13 \frac{\varkappa}{m_Q}.  
\label{eq:centermass3}
\end{align}

The mass splittings within a representation are caused by the
linear-order $m_{\mathrm{s}}$ contributions to the masses of the
baryon antidecapentaplet. The expressions of the $m_{\mathrm{s}}$ are
listed in Table~\ref{tab:1}.   
\setlength{\tabcolsep}{5pt}
\renewcommand{\arraystretch}{1.5}
\begin{table}[htp]
\caption{$m_{s}$ corrections to the masses of the baryon
  antidecapentaplet} 
\centering
\label{tab:1}
\begin{tabular}{l | c  c} 
\hline
$B$ & $M^{(1)}_{B,\bm{15}_{J=0}}$& $M^{(1)}_{B,\bm{15}_{J=1}}$ \\ 
\hline
$B_Q$ & $\dfrac{1}{4}\alpha+    \frac{5}{3} \beta  $ &
  $\frac{1}{8}\alpha+\frac{5}{3} \beta-\frac{1}{4}\gamma  $  \\  
$\Sigma_Q$ & $\frac{2}{3} \beta $  & $\frac{1}{12}\alpha+\frac{2}{3} 
\beta-\frac{1}{6}\gamma  $  \\ 
$\Lambda_Q$ & $\frac{1}{4}\alpha+  \frac{2}{3} \beta $ 
& $ \frac{2}{3} \beta                     $  \\ 
$\Xi_Q$ & $\frac{1}{8}\alpha -\frac{1}{3} \beta $  
& $-\frac{1}{12}\alpha-\frac{1}{3} \beta+\frac{1}{6}\gamma   $  \\ 
$\Xi_{Q}^{3/2}$ & $-\frac{1}{4}\alpha -\frac{1}{3} \beta $  
& $\frac{1}{24}\alpha-\frac{1}{3} \beta-\frac{1}{12}\gamma $  \\ 
$\Omega_Q$ & $ -\frac{4}{3} \beta $  
& $-\frac{1}{6}\alpha-\frac{4}{3} \beta+\frac{1}{3}\gamma   $  \\ 
\hline
\end{tabular}
\end{table}

\section{Result and Discussion \label{sec:3}}
The $\chi$QSM contains basically three parameters, i.e. the current
quark masses $m_0$ and $m_{\mathrm{s}}$, the cut-off masses for the
proper-time regularization of the quark loops, the pion decay constant
$f_\pi$, and the dynamical quark mass $M$.    
In Ref.~\cite{Blotz:1992pw}, it was shown how to fix them. $m_0$ and
the cut-off masses are fixed by reproducing the experimental data of
the pion mass $m_{\pi}=139.57$~MeV and the pion decay constant
$f_{\pi}=93$~MeV, respectively. The strange current quark mass can be
also determined by reproducing the kaon mass. However, we will fix it
to be $m_{\mathrm{s}}=180$ MeV by using the mass splittings of the
baryon octet. The dynamical quark mass is considered to
be a free parameter. However, it is determined to be $M=420$ MeV such
that the electric charge radius of the proton is reproduced. Though
the parameters $\varkappa/m_Q$ were introduced phenomenologically,
their values were already fixed by the splitting between the baryon
sextet representations with $J'=1/2$ and $J'=3/2$. Thus, we have no
more free parameter to fit in the present work.       

\begin{figure}[htp]
\centering
\includegraphics[scale=0.267]{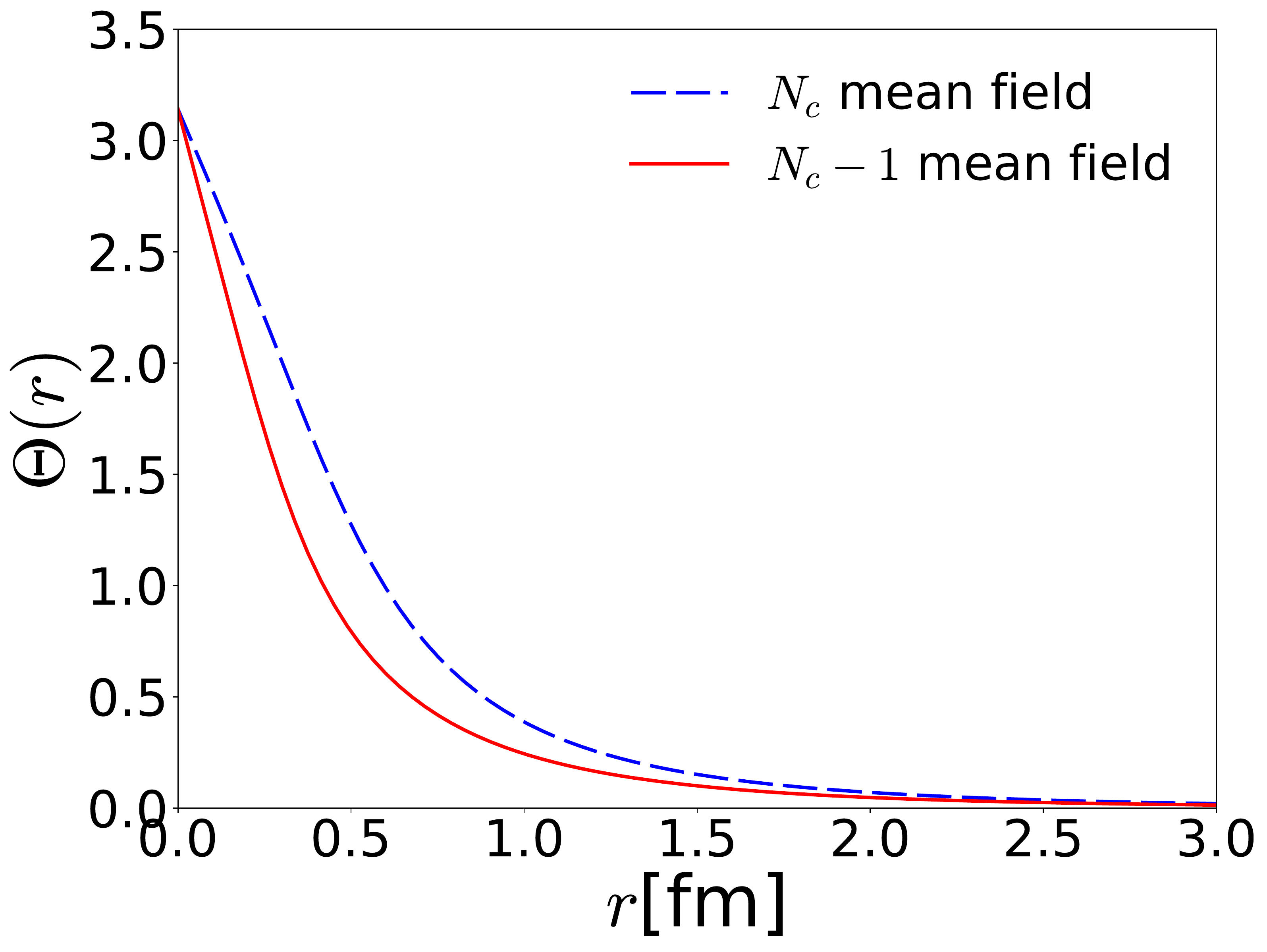}
\includegraphics[scale=0.267]{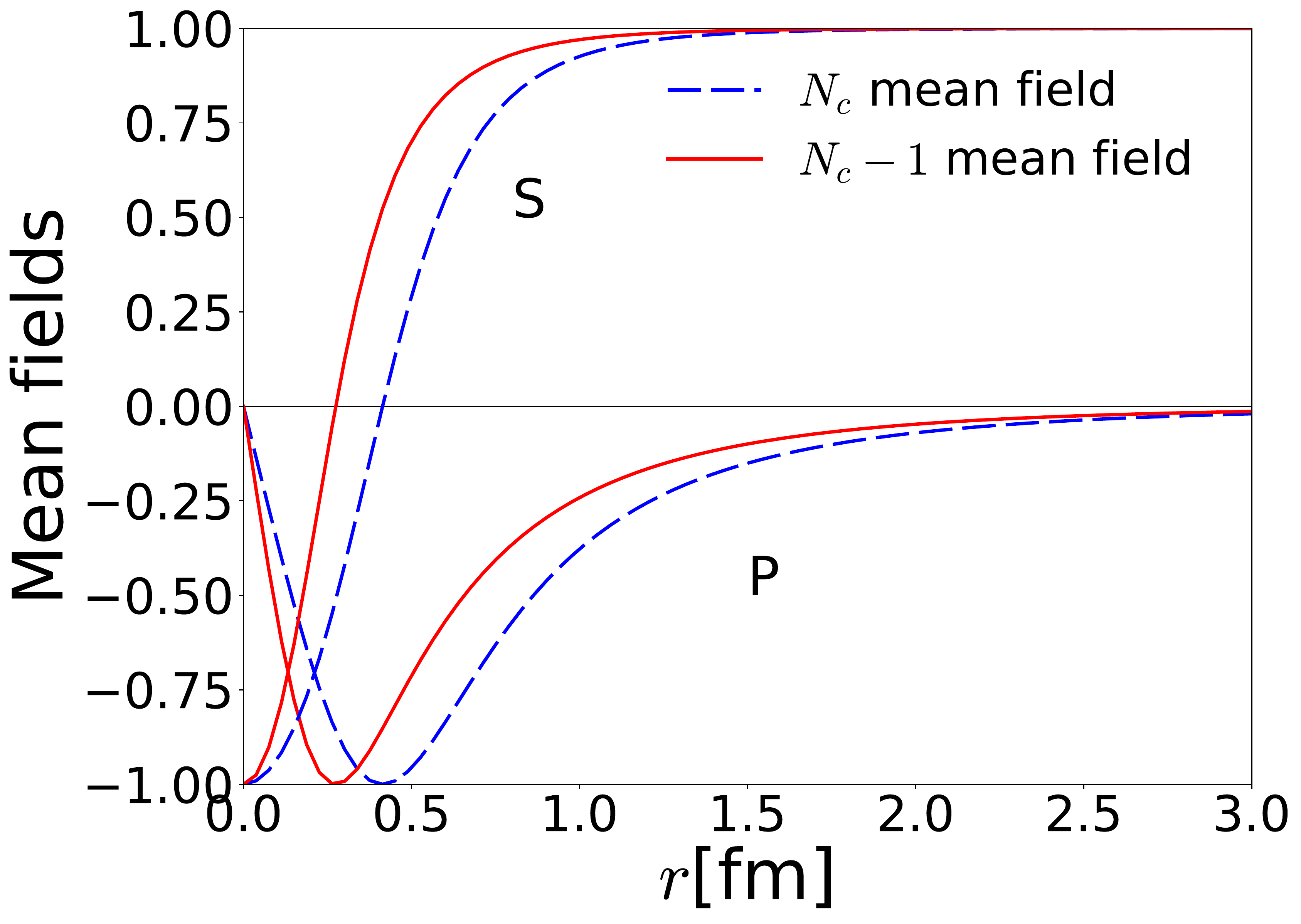}
\caption{The results of the self-consistent
  profile functions $\Theta(r)$ in the left panel and those of the 
  self-consistent scalar and pseudoscalar mean fields in the right
  panel. The dashed curves depict the results for the light baryons,
  whereas the solid ones draw those for the singly-heavy baryons. The
  value of the dynamical quark mass $M=420$ MeV is used.}  
\label{fig:1}
\end{figure}
To derive the profile function $\Theta(r)$, we first have to
solve the classical equation of motion~\eqref{eq:eq_motion}
self-consistently. Using a trial profile function, for which we use
either the linear profile function or the arctangent one, we solve the
one-body Dirac equation~\eqref{eq:one_enregy}, so that we obtain the
eigen-energies and eigenfunctions of the valence and sea
quarks. Inserting them into Eq.~\eqref{eq:densities} and solving
Eq.~\eqref{eq:eq_motion}, we find a new profile function. We repeat
this process until we obtain the classical energy, which converges
enough. This is nothing but a well-known Hartree approximation. 
In the left panel of Fig.~\ref{fig:1}, we draw the results of the
self-consistent profile functions for both the light and singly heavy
baryons with $M=420$ MeV used. The dashed curves exhibit that for the
light baryons and the solid ones show that for the singly heavy
baryons. As for the doubly heavy baryons, $M=420$ MeV is not strong
enough to find the minimum solution of the classical equation of
motion. As will be discussed later, a solution for the doubly heavy
baryons only appears when $M$ is larger than around 600 MeV. 
As shown in the left panel of Fig.~\ref{fig:1}, the profile function
for the singly heavy baryons shrinks from that for the light
baryons. Since the number of the light quarks inside a singly heavy
baryon is less than that inside a light baryon, the strength of the
pion mean fields is weakened. As a result, the size of the soliton is
also decreased. The soliton size is around 0.6 fm for the light
baryons with $N_c=3$, whereas it is around 0.4 fm for the singly heavy
baryons. In the right panel of Fig.~\ref{fig:1}, we draw the scalar
and pseudoscalar mean fields $S(r)$ and $P(r)$, which are defined in
Eq.~\eqref{eq:densities}.  As expected from the results of 
the profile functions, the scalar and pseudoscalar mean-field
densities are also shifted to the core of the soliton. 

We want to emphasize that the soliton for the $N_c-1$ valence quarks
is naturally the bosonic one, whereas that for the $N_c$ is the
fermionic one. This is a unique feature of the $\chi$QSM, which is
distinguished from any topological chiral soliton models, including
the Skyrme models, where the baryon number is identified with the
topological winding number or topological charge from the
Wess-Zumino term. In the Skyrme models, the integer winding number is
essential to have a finite energy for a stable topological soliton. On
the other hand, we do not need to have the integer winding number,
since the baryon number is constrained by the valence quarks. The
baryon number of the singly heavy baryon will be given by the $N_c-1$
light valence quarks together with a heavy quark. This is a
distinguished feature from a Skyrme model for heavy
baryons~\cite{Momen:1993ax}. We will discuss it in more detail later.  

\begin{figure}[htp]
\includegraphics[scale=0.4]{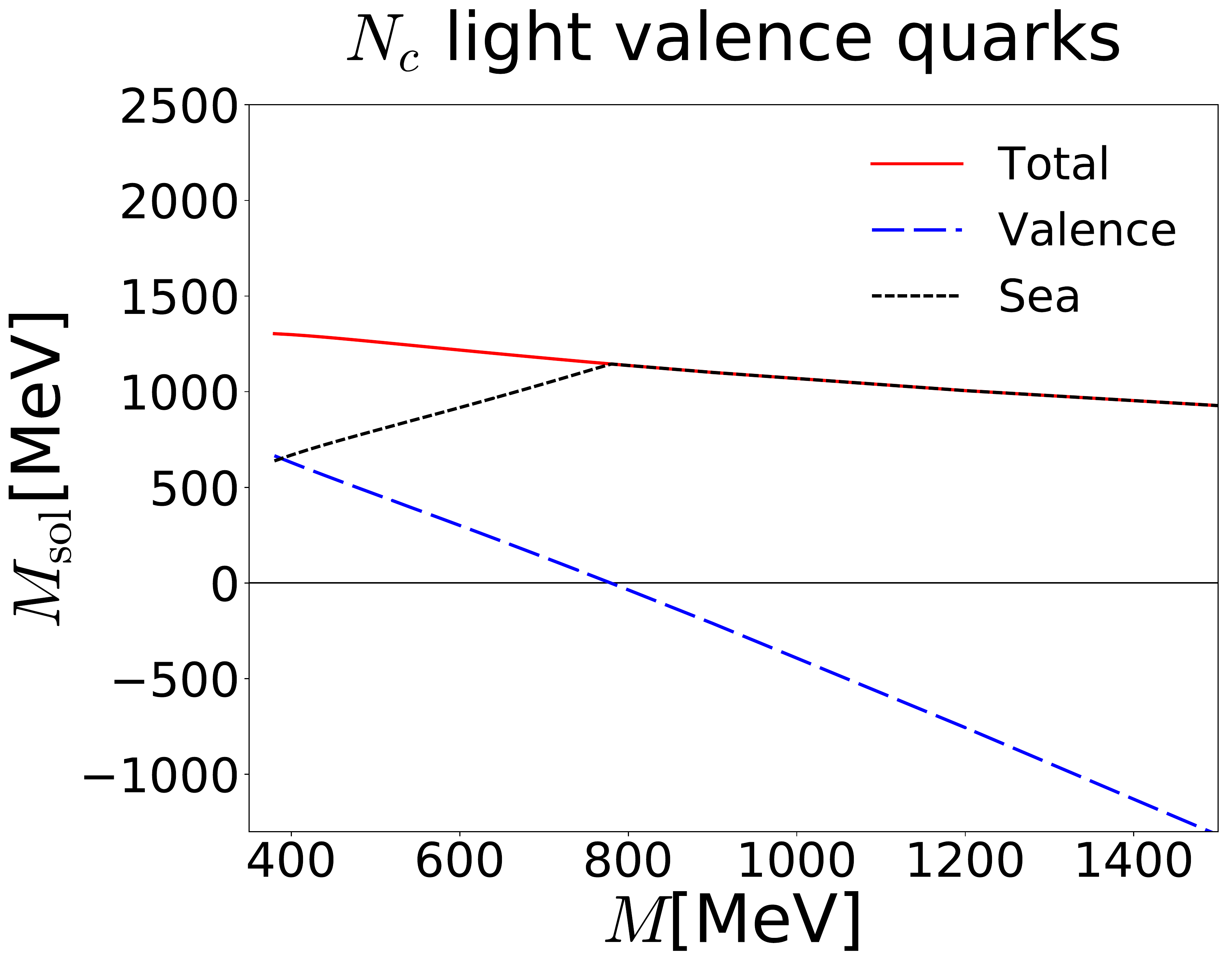}
\caption{Soliton mass as a function of the dynamical quark mass $M$
  for the $N_c$ mean field. The long-dashed line draws the
  valence-quark contribution, whereas the short-dashed one depicts the
  sea-quark contribution. The solid line represents the soliton mass.}    
\label{fig:2}
\end{figure}
In Fig.~\ref{fig:2}, we draw the soliton mass $M_{\mathrm{sol}}$
defined in Eq.~\eqref{eq:solnc} as a function of the dynamical quark
mass $M$. The results are the same as in
Ref.~\cite{Christov:1995vm}. When the $N_c$ valence quarks are
present, the mean-field solutions of the classical equation of the
motion exist when $M$ is larger than the critical mass
$M_{\mathrm{cr}}\approx 350$ MeV below which there is no
solution. Note that the dynamical quark mass plays a role of the
coupling between the quark and the pion. So, $M$ is smaller than
$M_{\mathrm{cr}}$, the interaction strength is not enough to bind the
$N_c$ valence quarks. As $M$ increases above $M_{\mathrm{cr}}$, the
valence-quark energy starts to decrease monotonically, whereas the
sea-quark energy increases. It indicates that as $M$ becomes larger,
the vacuum is polarized more strongly. This is very important to
stabilize the pion mean-field solution or the chiral soliton. Though
the valence- and sea-quark energies depend on $M$ rather sensitively,
the soliton mass decreases rather mildly, as $M$ increases. When the
value of $M$ is approximately 800 MeV, the valence level crosses the
line, below which the valence-quark energy turns negative. Then, the
soliton mass is only given by the sea-quark energy. In this case, the
baryon number is identified with the winding number. If we further
increase $M$, the valence level may dive into the negative Dirac
sea. This corresponds to the Skyrme picture of the baryon as a
topological soliton. This can be justified by the gradient
expansion. If $M >1$ GeV and the soliton rotates slowly,
then $\partial_k U/M$ becomes small. Hence, it can be used as an
expansion parameter to expand the baryon charge~\cite{Diakonov:1987ty}   
\begin{align}
B(U) &=  - \int \frac{d\omega}{2\pi} \mathrm{Tr}
  \left(\frac{1}{i\omega + h_{\mathrm{SU(2)}}(U)} - \frac{1}{i\omega +
  h_0}\right)    \cr
&= - \int \frac{d\omega}{2\pi} \mathrm{Tr}
  \left(\frac{h_{\mathrm{SU(2)}}(U)}{\omega^2 + h_{\mathrm{SU(2)}}^2(U)}
  - \frac{h_0}{\omega^2 +  h_0^2}\right), 
\end{align}
where $h_0$ is the Dirac Hamiltonian with the $U$ field turned off. 
$h_{\mathrm{SU(2)}}^2(U)$ and $h_0^2$ are written respectively as 
\begin{align}
h_{\mathrm{SU(2)}}^2 = -\partial_k^2 + M^2 + i M\gamma_k \partial_k
  U^{\gamma_5},\;\;\; 
h_0^2 = -\partial_k^2 + M^2.  
\end{align}
The leading term of the gradient term is obtained to be 
\begin{align}
B(U) = -\frac1{24\pi^2} \varepsilon_{ijk} \mathrm{Tr}
  \left[(U^\dagger \partial_i U) (U^\dagger \partial_j
  U)(U^\dagger \partial_k U) \right],
\end{align}
which is the well-known expression for the winding number in the
Skyrme model~\cite{Zahed:1986qz}. Thus, when $M$ becomes very large
($M>1$ GeV), one can see that the nucleon arises as a topological
soliton. We will see that in the case of the $N_c-1$ the mean-field
solution reveals a remarkable feature when $M$ increases. 

\begin{figure}[htp]
\includegraphics[scale=0.4]{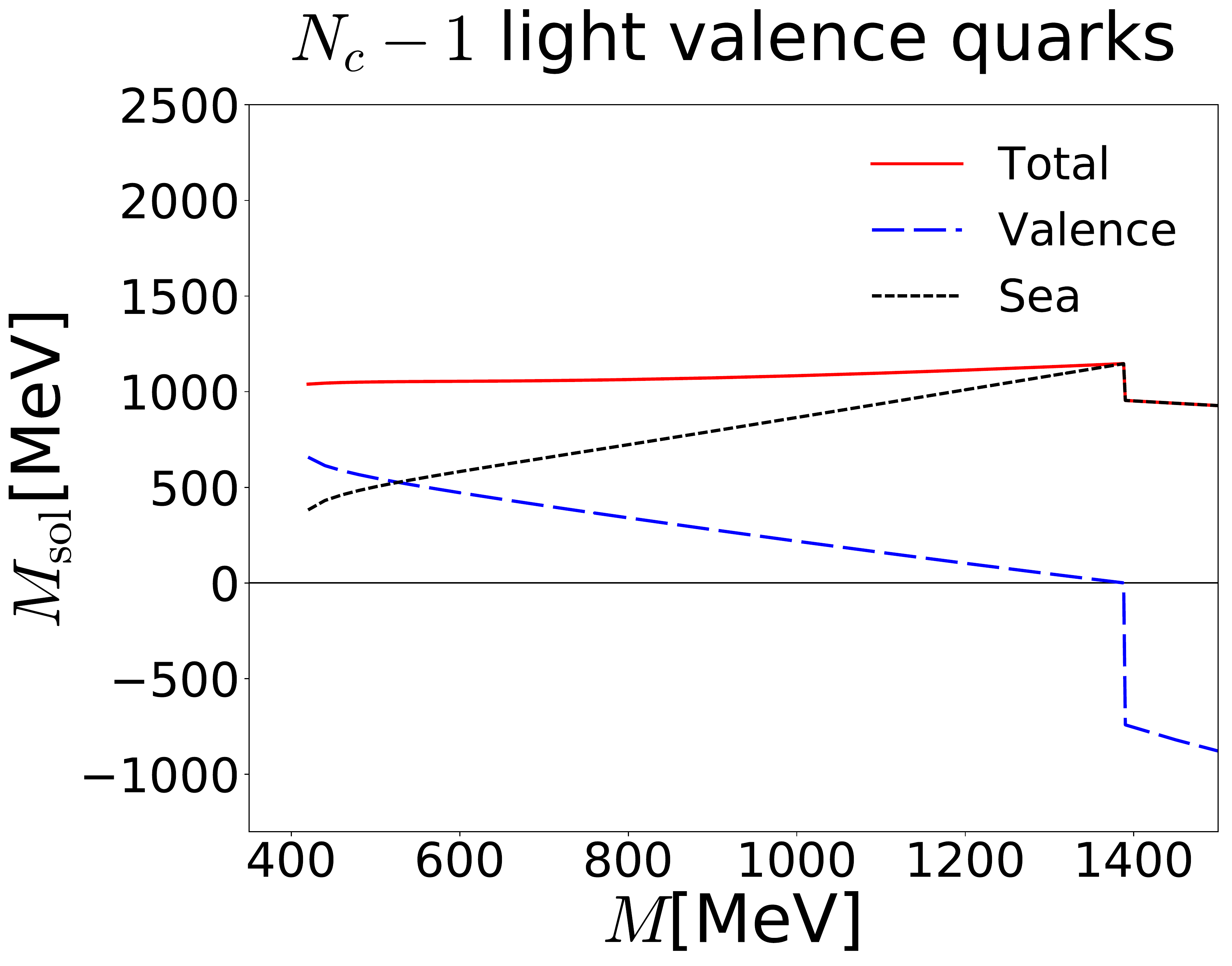}
\caption{Soliton mass as a function of the dynamical quark mass $M$
  for the $N_c-1$ mean fields. The long-dashed line draws the
  valence-quark contribution, whereas the short-dashed one depicts the
  sea-quark contribution. The solid line represents the soliton mass.}
\label{fig:3}
\end{figure}
Figure~\ref{fig:3} illustrates the soliton mass for the $N_c-1$ mean
fields, i.e., for the singly heavy baryons, as a function of $M$. Note
that in this case the soliton is the bosonic one consisting of two
valence quarks with $N_c=3$, as already mentioned previously. Since
the number of the valence quarks are reduced by one, we expect that
the Dirac-sea polarization becomes weaker than the case of the three
valence quarks. Indeed, the solitonic solution exists only when the
dynamical quark mass is larger than $M_{\mathrm{cr}}\approx 400$ MeV,
as shown in Fig.~\ref{fig:3}. The general tendency of the valence- and
sea-quark energies is similar to the case of the $N_c$ mean
field. However, the sea-quark energy increases faster than the rate
that the valence-quark one falls off as $M$ increases. As a result,
the soliton mass rises monotonically very mildly. However, when the
valence-quark level crosses the line at which the valence energy
vanishes, the soliton mass coincides exactly with the that of the
$N_c$ mean field. This can be understood by Eq.~\eqref{eq:solnc}. The
soliton mass comes solely from the sea-quark energy. Moreover, in this
case, the baryon number is acquired by the winding number as we
discussed previously. Thus, the present picture is reduced to the
Skyrme model for the singly heavy baryons~\cite{Momen:1993ax}, where
the singly heavy baryons were constructed by putting together the
topological soliton with the baryon number $B=1$ and a heavy
meson. However, we emphasize again that singly heavy baryons arise
from the $N_c-1$ bosonic soliton together with a heavy quark when the
plausible value of $M$ is used.   

\begin{figure}[htp]
\includegraphics[scale=0.4]{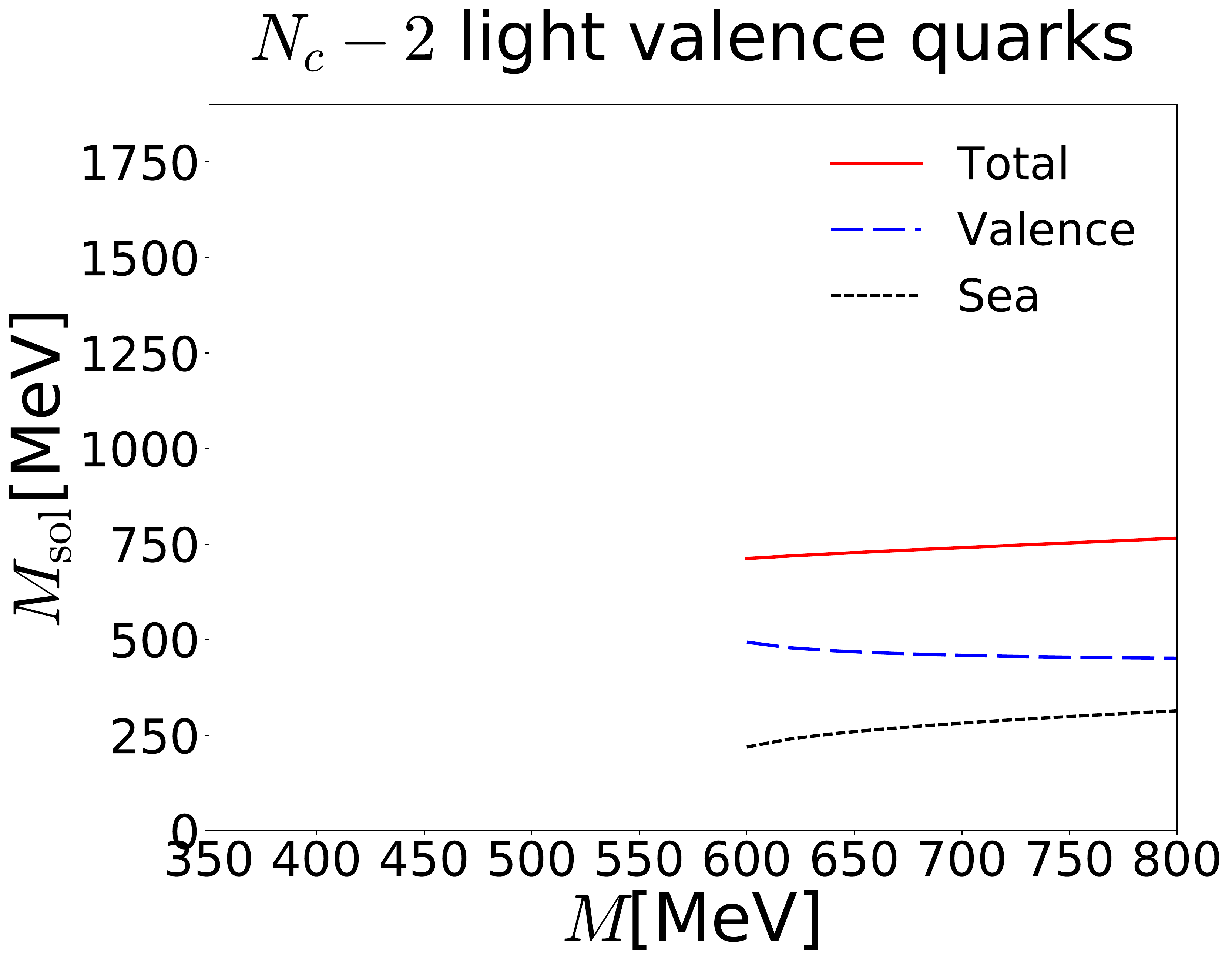}
\caption{Soliton mass as a function of the dynamical quark mass $M$
  for the $N_c-2$ mean fields. The long-dashed line draws the
  valence-quark contribution, whereas the short-dashed one depicts the
  sea-quark contribution. The solid line represents the soliton mass.}
\label{fig:4}
\end{figure}
In Fig.~\ref{fig:4} we draw the soliton mass for the $N_c-2$ mean
fields. In this case, the soliton is just a qualiton consisting of a 
single light valence quark. Unfortunately, the solitonic solution does
not exist with the dynamical quark mass $M=420$ MeV we adopt. The 
mean-field solution only appears when $M$ is larger than $M\approx
600$ MeV. Though $N_c-2$ is parametrically large, we have in practice
only one valence quark. Thus, we are not able to study the doubly
heavy baryons within the present approach. A single valence quark
is not enough to form a soliton with the proper value of the dynamical
quark mass. 

   \begin{table}[htp]
\setlength{\tabcolsep}{5pt}
\renewcommand{\arraystretch}{1.5}
  \caption{Results of the moments of inertia ($I_1$, $I_2$), the
    anomalous moments of inertia ($K_1$, $K_2$), the sigma $\pi N$
    terms, and the soliton mass for the 
    $N_c$ and $N_c-1$ mean fields. The dynamical quark mass $M=420$
    MeV is use. The second column lists the results
  for the $N_c$ mean field, whereas the fourth column shows those for
  the $N_c-1$ mean field. Note that $\overline{\Sigma}_{\pi N}$
  denotes the modified sigma $\pi N$ term. Its valence part is defined as
  $\overline{\Sigma}_{\pi N}^{\mathrm{val}} = (N_c-1) \Sigma_{\pi
    N}^{\mathrm{val}}/N_c$. Its sea part is the same as the original
sigma  $\pi N$ term.}  
  \label{tab:2}
\centering
\begin{tabular}{l c | c c} \hline
\hline
 & $N_{c}$ mean field & &$N_{c}-1$ mean field \\
\hline
$I_1^{\mathrm{val}}$[fm]  & 0.7923 & $I_1^{\mathrm{val}}$[fm] & 0.9591  \\
$I_1^{\mathrm{sea}}$[fm]  & 0.3137 & $I_1^{\mathrm{sea}}$[fm] & 0.1438  \\
$I_1$ \ [fm]   &1.1060& $I_1$ \ [fm]   & 1.1029  \\
\hline
$I_2^{\mathrm{val}}$[fm] & 0.3737 & $I_2^{\mathrm{val}}$[fm] & 0.4542 \\
$I_2^{\mathrm{sea}}$[fm] & 0.1551 & $I_2^{\mathrm{sea}}$[fm] & 0.0657 \\
$I_2$ \ [fm]  & 0.5288 & $I_2$ \ [fm]  & 0.5199 \\
\hline
$K_1^{\mathrm{val}}$[fm] & 0.4260 & $K_1^{\mathrm{val}}$[fm] & 0.6903 \\
$K_1^{\mathrm{sea}}$[fm] & 0.0009 & $K_1^{\mathrm{sea}}$[fm] & 0.0001 \\
$K_1$ \ [fm]  & 0.4269 & $K_1$ \ [fm]  & 0.6904 \\
\hline
$K_2^{\mathrm{val}}$[fm] & 0.2741 & $K_2^{\mathrm{val}}$[fm] & 0.3665 \\
$K_2^{\mathrm{sea}}$[fm] & -0.0019 & $K_2^{\mathrm{sea}}$[fm] & -0.0005 \\
$K_2$ \ [fm]  & 0.2722  & $K_2$ \ [fm]  & 0.3660 \\
\hline
${\Sigma}_{\pi N}^{\mathrm{val}}$  [MeV]& 11.07  &
 $\overline{\Sigma}_{\pi N}^{\mathrm{val}}$  [MeV]& 8.73 \\ 
${\Sigma}_{\pi N}^{\mathrm{sea}}$  [MeV]& 32.92  & 
$\overline{\Sigma}_{\pi N}^{\mathrm{sea}}$  [MeV]& 14.49 \\
${{\Sigma}}_{\pi N}$    [MeV] & 43.99  & $\overline{\Sigma}_{\pi N}$    
[MeV] & 23.22 \\
\hline
$M_{\mathrm{sol}}^{\mathrm{val}}$   [MeV] & 595  
& $M_{\mathrm{sol}}^{\mathrm{val}}$   [MeV] & 658 \\
$M_{\mathrm{sol}}^{\mathrm{sea}}$  [MeV] & 697  
& $M_{\mathrm{sol}}^{\mathrm{sea}}$  [MeV] & 382 \\
$M_{\mathrm{sol}}$  \  [MeV] & 1292  & $M_{\mathrm{sol}}$  \  
[MeV] & 1040 \\
\hline
\hline
\end{tabular}
\end{table}
In Table~\ref{tab:2}, we list in the second column the results of the 
moments of inertia, sigma $\pi N$ term, and the classical soliton mass
$M_{\mathrm{sol}}$ for the $N_c$ soliton. In the last column, the
results of the same quantities are listed for the $N_c-1$
soliton. Note that in the case of the $N_c-1$ mean 
fields, the valence parts of these quantities have $N_c-1$ factors
instead of $N_c$. What is interesting is that, naively thinking,
valence part of the moments of inertia with $N_c-1$ factors would
decrease in comparison with the results of those in the case of the
$N_c$ mean fields. However, the situation is more complicated and
dynamical. As shown in Table~\ref{tab:2}, the results of the valence
parts of the moments of inertia increase in comparison with the case
of the $N_c$ mean fields, whereas those of the sea parts of them
decrease by changing the mean fields from $N_c$ to $N_c-1$. The total
results of $I_1$ and $I_2$ with the $N_c-1$ mean fields are rather
similar to those with $N_c$ ones. On the other hand, the total results
of the anomalous moments of inertia $K_1$ and $K_2$ with the $N_c-2$
mean fields, which arise from the linear $m_s$ corrections, become
larger than those with $N_c$ mean fields. The $\overline{\Sigma}_{\pi
  N}$ and $M_{\mathrm{sol}}$ with $N_c-1$ mean fields are smaller than
those with $N_c$ ones.  

Since the $\Sigma_{\pi N}$ is proportional to the matrix element
$\langle n |\gamma_4 |n\rangle$ as shown in Eq.~\eqref{eq:sigmaT}, it
is easy to understand that the value of $\overline{\Sigma}_{\pi N}$
becomes smaller than that of $\Sigma_{\pi N}$. It is also natural that
$M_{\mathrm{sol}}$ with the $N_c-1$ mean fields turn out smaller than
that with the $N_c$ mean fields (see Eq.~\eqref{eq:solnc}). However,
when it comes to the moments of inertia, the denominators
$E_n-E_{\mathrm{val}}$ and $E_n-E_m$ with $N_c-1$ mean fields in
Eqs.~\eqref{eq:mom_exp1} and \eqref{eq:Kmom} become
smaller than those with $N_c$ mean fields, which brings about the
increase of the values of the moments of inertia.     
 
\begin{table}[htp]
\setlength{\tabcolsep}{5pt}
\renewcommand{\arraystretch}{1.5}
  \caption{Results of the dynamical parameters for the flavor SU(3)
    symmetry breaking. The strange current quark mass is taken to be
    $m_{s}=180$~MeV. The results are compared with the previous
    works~\cite{Kim:2018xlc, Yang:2016qdz}. In Ref.~\cite{Kim:2018xlc}
    all the results were computed without modifying the mean
    fields. In the last column, the values of
    $\delta_{\overline{\bm{3}}}$ and $\delta_{\bm{6}}$ were listed,
    which were determined by the experimental data.}    
  \label{tab:3}
\centering
\begin{tabular}{c c | c c c} 
\hline
\hline
[MeV] & This work & \cite{Kim:2018xlc}  & \cite{Yang:2016qdz} & 
Experiment~\cite{Tanabashi:2018oca} \\
\hline
$\alpha$ & -142.7 & -337.8 & -255.03$\pm$5.82 & - \\
$\beta$ & -126.7 & -80.6 & -140.04$\pm$3.20 & - \\
$\gamma$ & -28.10 & -39.3  & -101.08$\pm$2.33 & - \\
\hline
$\delta_{\overline{\bm{3}}}$ & -180.3 & -207.3  & -203.8$\pm$3.5
&$\sim$  -182.9   \\ 
$\delta_{\bm{6}}$ & -139.7 & -119.5  & -135.2$\pm$3.3 &$\sim$ -122.4 \\
\hline
\hline
\end{tabular}
\end{table}
In Table~\ref{tab:3}, we list the numerical results of the dynamical
parameters $\alpha$, $\beta$, and $\gamma$ in Eq.~\eqref{eq:alphaetc}
in comparison with the previous works. Those of
$\delta_{\overline{\bm{3}}}$ and $\delta_{\overline{\bm{3}}}$ in
Eq.~\eqref{eq:delta36} are also presented. The results are compared
with those of the previous work~\cite{Kim:2018xlc}, in which the $N_c$
mean fields were used without any modification, assuming that the mean 
fields would not be much changed. However, as shown in
Table~\ref{tab:3}, the results of $\alpha$, $\beta$, and $\gamma$ turn
out to be much different from those of Ref.~\cite{Kim:2018xlc}. What
is more interesting is that the results of
$\delta_{\overline{\bm{3}}}$ and $\delta_{\overline{\bm{3}}}$ are much
closer to those extracted from the experimental data, compared even
with those of Ref.~\cite{Yang:2016qdz}, in which all the dynamical
parameters were determined by the experimental data on the light
baryons. Note that since Ref.~\cite{Yang:2016qdz} performed a
``model-independent'' analysis, it is not possible to decompose the
valence-quark and sea-quark parts. To compensate this, an additional
scale factor was introduced in Ref.~\cite{Yang:2016qdz}. Thus, the
$N_c-1$ mean fields derived in the present work provide not only a
theoretically consistent method but also a phenomenologically better
description of the experimental data. 

\begin{table}[htp]
\setlength{\tabcolsep}{5pt}
\renewcommand{\arraystretch}{1.5}
\caption{Results of the masses of the ground-state charmed baryons in
  units of MeV in comparison with those of Refs.~\cite{Kim:2018xlc,
    Yang:2016qdz}. The results listed in the fifth
  column~\cite{Kim:2018xlc}$^{*}$ are rederived without fitting the
  center masses to the experimental data. For details, see the
  related text. The last column lists the experimental data.}   
\label{tab:4}
\centering 
\begin{tabular}{c c  c  c   c c  c}
\hline
\hline
${\cal{R}}^{Q}_{J'}$& $B_{c}$   & This work & \cite{Kim:2018xlc} &
 \cite{Kim:2018xlc}$^{*}$ & \cite{Yang:2016qdz} &
 Experiment~\cite{Tanabashi:2018oca}    \\   
\hline
$\overline{\bm{3}}^{c}_{1/2}$&$\Lambda_{c}$    
& 2278.4 & 2274.4 & 2225.4 & 2272.5$\pm$2.3 & 2286.5$\pm$0.1\\ 
$\overline{\bm{3}}^{c}_{1/2}$&$\Xi_{c}$        
& 2458.6 & 2481.5 & 2432.7 & 2476.3$\pm$1.2 & 2469.4$\pm$0.3 \\
$\bm{6}^{c}_{1/2}$           &$\Sigma_{c}$     
& 2438.6 & 2455.7 & 2472.0 & 2445.3$\pm$2.5 & 2453.5$\pm$0.1 \\
$\bm{6}^{c}_{1/2}$           &$\Xi'_{c}$       
& 2578.3 & 2575.2 & 2591.4 & 2580.5$\pm$1.6 & 2576.8$\pm$2.1 \\
$\bm{6}^{c}_{1/2}$           &$\Omega_{c}$     
& 2718.1 & 2694.6 & 2711.0 & 2715.7$\pm$4.5 & 2695.2$\pm$1.7 \\
$\bm{6}^{c}_{3/2}$           &$\Sigma^{*}_{c}$ 
& 2506.7 & 2523.9 & 2540.1 & 2513.4$\pm$2.3& 2518.1$\pm$0.8 \\
$\bm{6}^{c}_{3/2}$           &$\Xi^{*}_{c}$    
& 2646.4 & 2643.3 & 2659.6 & 2648.6$\pm$1.3& 2645.9$\pm$0.4 \\
$\bm{6}^{c}_{3/2}$           &$\Omega^{*}_{c}$ 
& 2786.2 & 2762.7 & 2779.1  & 2783.8$\pm$4.5& 2765.9$\pm$2.0 \\
\hline
\hline
\end{tabular}
\end{table}
In Table~\ref{tab:4}, we list the numerical results of the masses of
the lowest-lying charmed baryons, comparing them with those of
Ref.~\cite{Kim:2018xlc, Yang:2016qdz}. Note that in
Ref.~\cite{Kim:2018xlc} the center masses were rather different from
the experimental data. Thus, the relevant parameters had to be fitted
to the data. In the fifth column with~\cite{Kim:2018xlc}$^*$, we list
the results that were reevaluated without fitting the center masses.  
As can be observed by comparing the present results with those listed
in the fifth column, the $N_c-1$ pion mean fields describe the
experimental data far better than the $N_c$ mean fields. Even compared
with those of the model-independent analysis~\cite{Yang:2016qdz}, the
present results are quantitatively comparable with those of
Ref.~\cite{Yang:2016qdz}.      

\begin{table}[htp]
\setlength{\tabcolsep}{5pt}
\renewcommand{\arraystretch}{1.5}
\caption{Results of the masses of the ground-state bottom baryons in
  units of MeV in comparison with those of Refs.~\cite{Kim:2018xlc,
    Yang:2016qdz}. The results listed in the fifth
  column~\cite{Kim:2018xlc}$^{*}$ are rederived without fitting the
  center masses to the experimental data. For details, see the
  related text. The last column lists the experimental data.}
\label{tab:5}
\centering 
\begin{tabular}{c c  c  c   c c  c}
\hline
\hline
${\cal{R}}^{Q}_{J'}$& $B_{b}$   & This work & \cite{Kim:2018xlc} &
 \cite{Kim:2018xlc}$^{*}$ & \cite{Yang:2016qdz} & 
Experiment~\cite{Tanabashi:2018oca}    \\  
\hline
$\overline{\bm{3}}^{b}_{1/2}$&$\Lambda_{b}$    
& 5608.2 & 5602.7 & 5554.3 & 5599.3$\pm$2.4 & 5619.5$\pm$0.2 \\ 
$\overline{\bm{3}}^{b}_{1/2}$&$\Xi_{b}$        
& 5788.5 & 5809.9 & 5761.6 & 5803.1$\pm$1.2 & 5793.1$\pm$0.7 \\ 
$\bm{6}^{b}_{1/2}$           &$\Sigma_{b}$     
& 5800.3 & 5812.7 & 5832.7 & 5804.3$\pm$2.4 & 5813.4$\pm$1.3 \\ 
$\bm{6}^{b}_{1/2}$           &$\Xi'_{b}$       
& 5940.1 & 5932.1 & 5952.2 & 5939.5$\pm$1.5 & 5935.0$\pm$0.05 \\ 
$\bm{6}^{b}_{1/2}$           &$\Omega_{b}$     
& 6079.8 & 6051.6 & 6071.7 & 6074.7$\pm$4.5 & 6048.0$\pm$1.9 \\ 
$\bm{6}^{b}_{3/2}$           &$\Sigma^{*}_{b}$ 
& 5820.6 & 5834.7 & 5853.0 & 5824.6$\pm$2.3 & 5833.6$\pm$1.3 \\ 
$\bm{6}^{b}_{3/2}$           &$\Xi^{*}_{b}$    
& 5960.3 & 5954.2 & 5972.5 & 5959.8$\pm$1.2 & 5955.3$\pm$0.1 \\ 
$\bm{6}^{b}_{3/2}$           &$\Omega^{*}_{b}$ 
& 6100.1 & 6073.6 & 6092.0 & 6095.0$\pm$4.4 &       - \\  
\hline
\hline
\end{tabular}
\end{table}
In Table~\ref{tab:5}, we present the numerical results of the masses
of the lowest-lying bottom baryons. As in the case of the charmed
baryons, the results are in good agreement with the experimental
data. The present model predicts the $\Omega_b^*$ mass to be $6100.1$
MeV, which is slightly larger than that of
Ref.~\cite{Yang:2016qdz}. Again, this work reproduces 
quantitatively the data by far better than
Ref.~\cite{Kim:2018xlc}$^*$.  

\section{Summary and conclusions}
In the present work, we investigated how the pion mean fields
underwent the changes within the framework of the self-consistent
chiral quark-soliton model, when the number of the light valence
quarks is reduced from $N_c$ to $N_c-N_Q$. Starting from the baryon
correlation functions with $N_c-1$ valence quarks, we derived the
classical energy of the classical soliton. Having minimized the
energy, we found the equation of motion. Having solved it in a
self-consistent way, we obtained the numerical result of the profile
function for the classical soliton solution or the pion mean
fields. Compared with the profile function for the $N_c$ pion mean
fields, that for the $N_c-1$ mean fields shrinks to the core of the 
soliton. We also found that the solution for the $N_c-2$ mean fields
did not exist with the proper value of the dynamical quark mass. It
implies that the present scheme is not suitable for the description of
the doubly heavy baryons. The soliton mass was also investigated as a
function of the dynamical quark mass $M$. The solution exists when 
$M$ is approximately larger than 400 MeV. As $M$ increases, the
valence-quark contribution falls off slowly whereas the sea-quark
contribution increases also mildly. However, when $M$ reaches around
1390 MeV, where the valence-quark part vanishes, the sea-quark
contribution coincides exactly with that of the soliton mass with the
$N_c$ valence quarks. It indicates that when the valence energy
crosses the line at which it becomes zero the baryon number is
acquired by the winding number or the topological charge as in the
case of the Skyrme model. We want to emphasize that the soliton is
made of the $N_c-1$ valence quarks, i.e. a bosonic soliton with the
proper value of the dynamical quark mass, which is conceptually
different from any topological chiral soliton models including the
Skyrme model. 

The moments of inertia and anomalous moments of inertia
become larger than those with the $N_c$ mean fields. The changes from
the $N_c$ to $N_c-1$ valence quarks give rise to nontrivial effects on
these dynamical quantities of the classical soliton. 
The results of the masses of the charmed and bottom baryons show that
the modification of the pion mean fields reproduce the experimental
data much better than the previous work if one does not fit the center
masses of the heavy baryons. The $\Omega_b^*$ baryon is predicted to
be $M_{\Omega_b^*}=6100.1$ MeV, which is slightly larger than that
from the model-independent analysis. We also computed the masses of
the charmed and bottom baryon antidecapentaplet. In general, the
present work yield larger values of the masses in comparison with the
previous works except for the $B_c$ baryons. 

In conclusion, it is essential to consider the modification of the
pion mean fields when the number of the valence quarks is reduced from 
$N_c$ to $N_c-1$. The $N_c-1$ mean fields provide a better
understanding of the nature of the singly heavy baryons and a more
quantitative description of the masses of the charmed and bottom
baryons than the $N_c$ pion mean fields. Though the numbers of the
valence quarks $N_c$ and $N_c-1$ seem to be parametrically the same,
they yield different results in reality. The $N_c-1$ pion mean-field
solutions can be also used for the investigation of various
observables and form factors of the singly heavy baryons. The
corresponding works are under way. 

\section*{Acknowledgments}
The present work was supported by Basic Science Research Program
through the National Research Foundation of Korea funded by the
Ministry of Education, Science and Technology
(Grant-No. 2018R1A2B2001752 and 2018R1A5A1025563). 
J.-Y. Kim is partially supported by a DAAD doctoral scholarship.  

\appendix
\section{Moments of inertia\label{app:A}}
In this Appendix, we present all relevant formulae for the modified 
$\pi N$ sigma term, the moments of inertia $I_{1,2}$, $K_{1,2}$,
and $N_{1,2}$. The modified $\pi N$ sigma term is expressed as 
\begin{align}
\overline{\Sigma}_{\pi N}= \overline{\Sigma}^{\mathrm{val}}_{\pi
  N}+\Sigma^{\mathrm{sea}}_{\pi N},
\label{eq:app1}
\end{align}
where the valence and sea parts are written respectively as 
\begin{align}
\overline{\Sigma}^{\mathrm{val}}_{\pi N}  = m_0 (N_{c}-1) \langle
  \mathrm{val} | 
\gamma_{4}  | \mathrm{val} \rangle,\;\;\;
\Sigma^{\mathrm{sea}}_{\pi N}=\frac{m_0}{2}N_{c}\sum_{n}\langle n |
\gamma_{4}  | n \rangle \mathrm{sign}(E_{n}){{\cal{R}}_{{\Sigma}}} (E_n),
\label{eq:sigmaT}
\end{align}
where $\gamma_4$ denotes the Dirac $\gamma$ matrix in Euclidean space
\begin{align} 
\gamma_4 = 
  \begin{pmatrix}
    \bm{1} & 0 \\ 0 & -\bm{1}
  \end{pmatrix}.
\end{align}
The function $\mathcal{R}_\Sigma(E_n)$ stands for a regulator
\begin{align}
{\cal{R}}_{\Sigma}(E_n)=\frac{1}{\sqrt{\pi}} \int^{\infty}_{0} 
  \frac{du}{\sqrt{u}} e^{-u} \phi(u/E_n^2),
\end{align}
where $\phi(u)$~\cite{Blotz:1992pw} represents a cutoff function
defined by  
\begin{align}
\phi(u) = c \theta(u-1/\Lambda_1^2) + (1-c)\theta(u-1/\Lambda_2^2).  
\end{align}
The free parameters $\Lambda_1$, $\Lambda_2$, and $c$ are determined
by reproducing the pion decay constant $f_\pi=93$ MeV and the pion
mass $m_\pi=139$ MeV in the mesonic sector. The corresponding
numerical values are explicitly given as $\Lambda_1=381.15$ MeV,
$\Lambda_2=1428.00$ MeV, and $c=0.7276$.  

The moment of inertia tensor $I_{ab}$ is given as follows: 
\begin{align}
I_{\mathrm{ab}} =
  I^{\mathrm{val}}_{\mathrm{ab}}+I^{\mathrm{sea}}_{\mathrm{ab}}, 
\label{eq:momIi}
\end{align}
where 
\begin{align}
I^{\mathrm{val}}_{\mathrm{ab}} &=
  \frac{(N_{c}-1)}{2}\sum_{\mathrm{val,n \ne val}}\frac{\langle n |
  \lambda_{a} | \mathrm{val} \rangle \langle \mathrm{val} |
  \lambda_{b} | n \rangle}{E_{n}-E_{\mathrm{val}}},\cr 
I^{\mathrm{sea}}_{\mathrm{ab}}& = \frac{N_{c}}{4}\sum_{m,n}\langle
  n | \lambda_{a} | m \rangle \langle m | \lambda_{b} | n \rangle
  {\cal{R}}_{I}(E_{n},E_{m}), 
\label{eq:mom_exp1}
\end{align}
with the different regulator $R_{I}(E_n,\,E_m)$ 
\begin{align}
{\cal{R}}_{I}(E_{n},E_{m}) = \frac{1}{2\sqrt{\pi}}
  \int_{0}^{\infty} \frac{du}{\sqrt{u}} \phi(u)    
\left [ \frac{e^{-u E^{2}_{n}}-e^{-u E^{2}_{m}}}{u(E^{2}_{m}-E^{2}_{n})} 
- \frac{E_{n}e^{-u E^{2}_{n}}+E_{m}e^{-u E^{2}_{m}}}{E_{m}+E_{n}}
  \right ]. 
\end{align}
$\lambda_a$ in Eq.~(\ref{eq:mom_exp1}) stand for the Gell-Mann
matrices for flavor SU(3) group, satisfying the relations 
$\mathrm{tr}(\lambda_a\lambda_b) = 2\delta_{ab}$ and
$[\lambda_a,\,\lambda_b]=2if_{abc} \lambda_c$, 
$a=1,\cdots,8$. The moments of inertia $I_1$ and $I_2$ are defined by  
\begin{align}
I_{ab} \equiv \left \lbrace \begin{array}{c l}
  I_{1}\delta_{ab} & a,b=1,2,3 \\
  I_{2}\delta_{ab} & a,b=4,5,6,7\\
  0               & a,b=8
\end{array} \right. .
\end{align}
Similarly, the anomalous moments of inertia tensor is written as 
\begin{align}
K_{\mathrm{ab}}=K^{\mathrm{val}}_{\mathrm{ab}}+K^{\mathrm{sea}}_{\mathrm{ab}}, 
\end{align}
where
\begin{align}
K_{\mathrm{ab}}^{\mathrm{val}}&=\frac{(N_{c}-1)}{2} \sum_{\mathrm{val,n
  \ne val}}\frac{\langle n | \lambda_{a} | \mathrm{val} \rangle
  \langle \mathrm{val} | \lambda_{b} \gamma_{4} | n
  \rangle}{E_{n}-E_{\mathrm{val}}}, \cr
K_{\mathrm{ab}}^{\mathrm{sea}}&=\frac{N_{c}}{8}\sum_{m,n}\langle
  n | \lambda_{a} | m \rangle \langle m | \gamma_{4}\lambda_{b} | n
  \rangle \frac{\mathrm{sign}(E_{n})-\mathrm{sign}(E_{m}) }
  {E_{n}-E_{m}}.  
\label{eq:Kmom}
\end{align}
The anomalous moments of inertia $K_1$ and $K_2$ are defined by 
\begin{align}
K_{ab} \equiv \left \{ \begin{array}{c l}
  K_{1}\delta_{ab} & a,b=1,2,3 \\
  K_{2}\delta_{ab} & a,b=4,5,6,7\\
  0                & a,b=8
\end{array}   \right. .
\end{align}

\section{Matrix elements of the SU(3) Wigner $D$
  functions\label{app:b}} 
In Appendix~\ref{app:b}, we tabulate all relevant matrix elements of the
SU(3) Wigner $D$ functions in each representation. 
\begin{table}[htp]
\caption{Matrix elements of the SU(3) Wigner $D$ functions
  $D_{88}^{(8)}$ and $D_{8i}^{(8)} J_i$.}
\begin{tabular}{c c c c c c }
\hline
\hline 
&$ {\mathcal{R}}$ &T& Y & $ \langle {\cal{R}} Y T J|D^{(8)}_{88}|
{\mathcal{R}} Y T J \rangle$ &$\langle {\cal{R}} Y T J|D^{(8)}_{8i}J_{i}|
{\mathcal{R}} Y T J \rangle$   \\ 
 \hline
$B_{Q}$&\multirow{ 6}{*}{$\overline{\bf{15}}~(J=0)$} & $1/2$ & $5/3$ &   $1/4$ & 0 \\ 
$\Sigma_{Q}$& &$1$ & $2/3$   &  $0$ & 0 \\ 
$\Lambda_{Q}$& &$0$ &$ 2/3$   &  $1/4$ & 0 \\ 
$\Xi_{Q}$& &$1/2$ &$-1/3$   &  $1/8$ & 0 \\ 
$\Xi^{3/2}_{Q}$& &$3/2$ &$ -1/3$   &  $-1/4$ & 0 \\ 
$\Omega_{Q}$& & $1$ &$ -4/3$   &  $0$ & 0 \\ 
\hline
$B_{Q}$&\multirow{ 6}{*}{$\overline{\bf{15}}~(J=1)$} & $1/2$ & $5/3$ &   $1/8$ & $-1/4$ \\ 
$\Sigma_{Q}$& &$1$ & $2/3$   &  $1/12$ & $-1/6$ \\ 
$\Lambda_{Q}$& &$0$ &$ 2/3$   &  $0$ & $0$ \\ 
$\Xi_{Q}$& &$1/2$ &$-1/3$   &  $-1/12$ & $1/6$ \\ 
$\Xi^{3/2}_{Q}$& &$3/2$ &$ -1/3$   &  $1/24$ & $-1/12$ \\ 
$\Omega_{Q}$& & $1$ &$ -4/3$   &  $-1/6$ & $1/3$ \\ 
\hline
\hline
\end{tabular}
\label{tab:Dfuncion2}
\end{table}
\section{Mass spectra of the baryon antidecapentaplet\label{app:c}}  
\begin{table}[htp]
\setlength{\tabcolsep}{5pt}
\renewcommand{\arraystretch}{1.5}
\caption{Results of the masses of the charmed baryon antidecapentaplet
  in units of MeV in comparison with those of
  Ref.~\cite{Kim:2017jpx}. The fourth column lists the results 
  obtained by using the $N_c$ pion mean fields~\cite{Kim:2018xlc}.}
\label{tab:6}
\centering 
\begin{tabular}{c c c c c c c c}
\hline \hline 
$\mathcal{R}^{Q}_{J'}$ & $B_{Q}$   & This work &
 \cite{Kim:2018xlc}$^{*}$ & \cite{Kim:2017jpx} \\  
\hline
$\overline{\bm{15}}^{c}_{1/2}(J=0)$ & $B_c$           & 2909 & 3209 & - \\
$\overline{\bm{15}}^{c}_{1/2}(J=0)$ & $\Sigma_c$      & 3072 & 3374 & - \\
$\overline{\bm{15}}^{c}_{1/2}(J=0)$ & $\Lambda_c$     & 3036 & 3290 & - \\
$\overline{\bm{15}}^{c}_{1/2}(J=0)$ & $\Xi_c$         & 3181 & 3413 & - \\
$\overline{\bm{15}}^{c}_{1/2}(J=0)$ & $\Xi_{c}^{3/2}$ & 3234 & 3539 & - \\
$\overline{\bm{15}}^{c}_{1/2}(J=0)$ & $\Omega_c$      & 3325 & 3536 & - \\
$\overline{\bm{15}}^{c}_{1/2}(J=1)$ & $B_c$           & 2682 & 2952 & 2685 \\
$\overline{\bm{15}}^{c}_{1/2}(J=1)$ & $\Sigma_c$      & 2819 & 3053 & 2808 \\
$\overline{\bm{15}}^{c}_{1/2}(J=1)$ & $\Lambda_c$     & 2826 & 3075 & 2806 \\
$\overline{\bm{15}}^{c}_{1/2}(J=1)$ & $\Xi_c$         & 2960 & 3177 & 2928 \\
$\overline{\bm{15}}^{c}_{1/2}(J=1)$ & $\Xi_{c}^{3/2}$ & 2949 & 3145 & 2931 \\
$\overline{\bm{15}}^{c}_{1/2}(J=1)$ & $\Omega_c$      & 3094 & 3279 & 3050 \\
$\overline{\bm{15}}^{c}_{3/2}(J=1)$ & $B_c$           & 2750 & 3020 & 2754 \\
$\overline{\bm{15}}^{c}_{3/2}(J=1)$ & $\Sigma_c$      & 2887 & 3121 & 2877\\
$\overline{\bm{15}}^{c}_{3/2}(J=1)$ & $\Lambda_c$     & 2894 & 3143 & 2875\\
$\overline{\bm{15}}^{c}_{3/2}(J=1)$ & $\Xi_c$         & 3028 & 3245 & 2997\\
$\overline{\bm{15}}^{c}_{3/2}(J=1)$ & $\Xi_{c}^{3/2}$ & 3017 & 3213 & 3000\\
$\overline{\bm{15}}^{c}_{3/2}(J=1)$ & $\Omega_c$      & 3162 & 3347 & 3119\\
\hline \hline
\end{tabular}
\end{table}
\begin{table}[htp]
\setlength{\tabcolsep}{5pt}
\renewcommand{\arraystretch}{1.5}
\caption{Results of the masses of the charmed baryon antidecapentaplet
  in units of MeV in comparison with those of
  Ref.~\cite{Kim:2017jpx}. The fourth column lists the results 
  obtained by using the $N_c$ pion mean fields~\cite{Kim:2018xlc}.}
\label{tab:7}
\centering 
\begin{tabular}{c c c c c c c c}
\hline \hline
 $\mathcal{R}^{Q}_{J'}$ & $B_{Q}$   & This work & \cite{Kim:2018xlc}$^{*}$ & \cite{Kim:2017jpx} \\ 
\hline
$\overline{\bm{15}}^{b}_{1/2}(J=0)$ & $B_b$          & 6239 & 6511 & - \\
$\overline{\bm{15}}^{b}_{1/2}(J=0)$ & $\Sigma_b$     & 6402 & 6674 & - \\
$\overline{\bm{15}}^{b}_{1/2}(J=0)$ & $\Lambda_b$    & 6366 & 6638 & - \\
$\overline{\bm{15}}^{b}_{1/2}(J=0)$ & $\Xi_b$        & 6511 & 6782 & - \\
$\overline{\bm{15}}^{b}_{1/2}(J=0)$ & $\Xi_{b}^{3/2}$& 6564 & 6836 & - \\
$\overline{\bm{15}}^{b}_{1/2}(J=0)$ & $\Omega_b$.    & 6655 & 6927 & - \\
$\overline{\bm{15}}^{b}_{1/2}(J=1)$ & $B_b$          & 6043 & 6261 & 6044 \\
$\overline{\bm{15}}^{b}_{1/2}(J=1)$ & $\Sigma_b$     & 6181 & 6399 & 6167 \\
$\overline{\bm{15}}^{b}_{1/2}(J=1)$ & $\Lambda_b$    & 6188 & 6406 & 6165 \\
$\overline{\bm{15}}^{b}_{1/2}(J=1)$ & $\Xi_b$        & 6322 & 6540 & 6287 \\
$\overline{\bm{15}}^{b}_{1/2}(J=1)$ & $\Xi_{b}^{3/2}$& 6311 & 6529 & 6290 \\
$\overline{\bm{15}}^{b}_{1/2}(J=1)$ & $\Omega_b$.    & 6456 & 6674 & 6409 \\
$\overline{\bm{15}}^{b}_{3/2}(J=1)$ & $B_b$          & 6064 & 6281 & 6065 \\
$\overline{\bm{15}}^{b}_{3/2}(J=1)$ & $\Sigma_b$     & 6201 & 6419 & 6188 \\
$\overline{\bm{15}}^{b}_{3/2}(J=1)$ & $\Lambda_b$    & 6208 & 6426 & 6186 \\
$\overline{\bm{15}}^{b}_{3/2}(J=1)$ & $\Xi_b$        & 6342 & 6560 & 6308 \\
$\overline{\bm{15}}^{b}_{3/2}(J=1)$ & $\Xi_{b}^{3/2}$& 6331 & 6549 & 6311 \\
$\overline{\bm{15}}^{b}_{3/2}(J=1)$ & $\Omega_b$.    & 6476 & 6694 & 6430 \\
\hline \hline
\end{tabular}
\end{table}
In Appendix~\ref{app:c}, we present the results for the masses of the
baryon antidecapentaplet for completeness. 
In Ref.~\cite{Kim:2017jpx}, the two of the newly found
$\Omega_c$'s~\cite{Aaij:2017nav} were interpreted as the $\Omega_c$'s
that belong to the baryon antidecapentaplet.  As shown previously, the
baryon antitriplet and sextet naturally arise as the representations
of the rotational excitations with $Y'=2/3$. The next allowed
representation is the baryon antidecapentaplet ($\overline{\bm{15}}$)  
with $Y'=2/3$. The valence quark content of the baryon decapentaplet
is $Qqqq\bar{q}$, where $Q$ and $q$ denote the heavy and light quarks
respectively. $\bar{q}$ stands for the anti-light quark. So, the
members of the $\overline{\bm{15}}$-plet are the pentaquark baryons
including one heavy quark. Coupling the soliton spins 0 and 1 to the
heavy-quark spin 1/2, we find three different spin representations in
the baryon antidecapentaplet with $1/2$ and $(1/2,\,3/2)$. As shown in
Eqs.~\eqref{eq:center_mass2} and \eqref{eq:centermass3},
the values of $M_{B,\overline{\bm{15}}_{1/2},J=0}^Q$ are larger
than those of $M_{B,\overline{\bm{15}}_{1/2},J=1}^Q$ and
$M_{B,\overline{\bm{15}}_{3/2},J=1}^Q$. 

Table~\ref{tab:6} lists the numerical results of the masses of the
charmed baryon antidecapentaplet. As mentioned previously,
the results of $M_{B,\overline{\bm{15}}_{1/2},J=0}^c$ are larger than
the corresponding ones of $M_{B,\overline{\bm{15}}_{1/2},J=1}^Q$ and
$M_{B,\overline{\bm{15}}_{3/2},J=1}^Q$. Though there are no
experimental data on them, it is of great interest to consider the
masses of $\Omega_c$'s, in particular, when the soliton spin is
$J=1$. In Ref.~\cite{Kim:2017jpx}, $\Omega_c(3050)$ and
$\Omega_c(3119)$ were interpreted as possible pentaquark states that
belong to the $\overline{\bm{15}}$-plet. In the present work, we
obtain $M_{\Omega_c, 1/2}=3094$ MeV and $M_{\Omega_c, 3/2}=3162$,
which are somewhat larger than those of Ref.~\cite{Kim:2017jpx}. In
Table~\ref{tab:7}, we list the results of the masses of the bottom 
baryon antidecapentaplet. In general, the present results are again
larger than those predicted by Ref.~\cite{Kim:2017jpx}.



\begin{thebibliography}{99}
\bibitem{Witten:1979kh} 
  E.~Witten,
  Nucl.\ Phys.\ B {\bf 160}, 57 (1979).

\bibitem{Witten:1983}
  E.~Witten,
  Nucl.\ Phys.\ B {\bf 223}, 422 (1983).

\bibitem{Witten:1983tx}
  E.~Witten,
  Nucl.\ Phys.\ B {\bf 223}, 433 (1983).

\bibitem{Diakonov:1987ty} 
  D.~Diakonov, V.~Y.~Petrov and P.~V.~Pobylitsa,
  Nucl.\ Phys.\ B {\bf 306} (1988) 809.
  
\bibitem{Christov:1995vm} 
  C.~V.~Christov, A.~Blotz, H.-Ch.~Kim, P.~Pobylitsa, T.~Watabe,
  T.~Meissner, E.~Ruiz Arriola and K.~Goeke, 
  Prog.\ Part.\ Nucl.\ Phys.\  {\bf 37}, 91 (1996).

\bibitem{Diakonov:1997sj} 
  D.~Diakonov,
  In *Peniscola 1997, Advanced school on non-perturbative quantum
  field physics* 1-55 [hep-ph/9802298].
  
\bibitem{Blotz:1992pw} 
  A.~Blotz, D.~Diakonov, K.~Goeke, N.~W.~Park, V.~Petrov and
  P.~V.~Pobylitsa, 
  Nucl.\ Phys.\ A {\bf 555}, 765 (1993).

\bibitem{Diakonov:2010tf} 
  D.~Diakonov, 
  arXiv:1003.2157 [hep-ph].
  
\bibitem{Yang:2016qdz} 
  Gh.-S.~Yang, H.-Ch.~Kim, M.~V.~Polyakov, and M.~Prasza{\l}owicz,
  Phys.\ Rev.\ D {\bf 94}, 071502 (2016).
  
\bibitem{Kim:2018xlc} 
  J.~Y.~Kim, H.-Ch.~Kim and G.~S.~Yang,
  Phys.\ Rev.\ D {\bf 98}, no. 5, 054004 (2018).

\bibitem{Yang:2018uoj} 
  G.~S.~Yang and H.-Ch.~Kim,
  Phys.\ Lett.\ B {\bf 781}, 601 (2018).

\bibitem{Kim:2018nqf} 
  J.~Y.~Kim and H.-Ch.~Kim,
  Phys.\ Rev.\ D {\bf 97}, 114009 (2018).

\bibitem{Aaij:2017nav} 
  R.~Aaij {\it et al.} [LHCb Collaboration],
  Phys.\ Rev.\ Lett.\  {\bf 118}, 182001 (2017).

\bibitem{Yelton:2018mag} 
  J.~Yelton {\it et al.} [Belle Collaboration],
  Phys.\ Rev.\ Lett.\  {\bf 121} 052003 (2018).
  
\bibitem{Kim:2017jpx} 
  H.-Ch.~Kim, M.~V.~Polyakov, and M.~Prasza{\l}owicz,
  Phys.\ Rev.\ D {\bf 96}, 014009 (2017);  {\bf 96}, 039902(E) (2017).

\bibitem{Kim:2017khv} 
  H.-Ch.~Kim, M.~V.~Polyakov, M.~Praszalowicz and G.~S.~Yang,
  Phys.\ Rev.\ D {\bf 96}, 094021 (2017)
  Erratum: [Phys.\ Rev.\ D {\bf 97}, 039901 (2018)].
  
\bibitem{Kim:2018cxv} 
  H.-Ch.~Kim,
  J.\ Korean Phys.\ Soc.\  {\bf 73}, no. 2, 165 (2018)
  [arXiv:1804.04393 [hep-ph]].

\bibitem{Polyakov:2018zvc} 
  M.~V.~Polyakov and P.~Schweitzer,
  Int.\ J.\ Mod.\ Phys.\ A {\bf 33}, 1830025 (2018).

\bibitem{Kim} J.-Y. Kim, H.-D. Son, M. V. Polyakov, and H.-Ch. Kim, in
  prepation (2020).

\bibitem{Tanabashi:2018oca} 
  M.~Tanabashi {\it et al.} [Particle Data Group],
  Phys.\ Rev.\ D {\bf 98}, no. 3, 030001 (2018).
\bibitem{Momen:1993ax} 
  A.~Momen, J.~Schechter and A.~Subbaraman,
  Phys.\ Rev.\ D {\bf 49}, 5970 (1994).

\bibitem{Zahed:1986qz} 
  I.~Zahed and G.~E.~Brown,
  Phys.\ Rept.\  {\bf 142}, 1 (1986).
\end{thebibliography}
\end{document}